\documentclass[preprint,12pt]{elsarticle}

\usepackage{color,amsmath,amssymb,graphicx,latexsym,subfigure}
\usepackage{threeparttable}

\begin{document}

\begin{frontmatter}
\title{Millisecond pulsar interpretation of the Galactic center gamma-ray 
excess}

\author{Qiang Yuan$^{a,b}$, Bing Zhang$^b$}

\address[a]{Key Laboratory of Particle Astrophysics,
Institute of High Energy Physics, Chinese Academy of Science,
Beijing 100049, P.R.China}
\address[b]{Department of Physics and Astronomy, University of Nevada Las Vegas,
NV 89154, USA}

\begin{abstract}

It was found in the Fermi-LAT data that there is an extended $\gamma$-ray 
excess in the Galactic center region. The proposed sources to be
responsible for the excess include dark matter annihilation or an
astrophysical alternative from a population of millisecond pulsars
(MSPs). Whether or not the MSP scenario can explain the data 
self-consistently has very important implications for the detection of
particle dark matter, which is however, subject to debate in the literature.
In this work we study the MSP scenario in detail, based on the detected 
properties of the MSPs by Fermi-LAT. We construct a model of Milky Way
disk-component MSPs which can reproduce the $\gamma$-ray properties of the 
observed Fermi-LAT MSPs, and derive the intrinsic luminosity function of the 
MSPs. The model is then applied to a bulge population of MSPs. We find that 
the extended $\gamma$-ray excess can be well explained by the bulge MSPs
without violating the detectable flux distribution of MSPs by Fermi-LAT.
The spatial distribution of the bulge MSPs as implied by the distribution
of low mass X-ray binaries follows a $r^{-2.4}$ profile, which is also 
consistent with the $\gamma$-ray excess data. We conclude that the MSP 
model can explain the Galactic center $\gamma$-ray excess self-consistently, 
satisfying all the current observational constraints.

\end{abstract}

\end{frontmatter}

\section{Introduction}

It has been reported that there is an extended $\gamma$-ray excess in the
Galactic center (GC) region in the Fermi Large Area Telescope (Fermi-LAT) 
data \cite{2009arXiv0910.2998G,2009arXiv0912.3828V,2011PhLB..697..412H,
2011PhLB..705..165B,2012PhRvD..86h3511A,2013PhRvD..88h3521G,
2014arXiv1402.4090A,2014arXiv1402.6703D}. The spatial distribution of the 
extended excess follows the square of a generalized Navarro-Frenk-White 
(gNFW, \cite{1997ApJ...490..493N,1996MNRAS.278..488Z}) profile with inner 
slope $\gamma\approx1.2$, and the $\gamma$-ray spectrum can be fitted with 
an exponential cutoff power-law or a log-parabolic form
\cite{2011PhLB..697..412H,2012PhRvD..86h3511A,2013PhRvD..88h3521G}.
The spatial extension of the excess is rather large. Daylan et al.
found that up to $12^{\circ}$ away from the GC the excess is still remarkable
\cite{2014arXiv1402.6703D}. The analysis of the spatial variation 
of the $\gamma$-ray emission from the Fermi bubbles \cite{2010ApJ...724.1044S} 
showed that there might also be an extra component overlapping on the bubble 
emission, which follows the same projected gNFW$^2$ distribution of the GC 
excess \cite{2013PDU.....2..118H,2013arXiv1307.6862H,2014arXiv1402.6703D}.
This means the excess may exist at even larger scales.

The origin of this excess is still unclear, and the proposed sources
include dark matter (DM) annihilation \cite{2011PhRvD..84l3005H,
2011JHEP...05..026M,2011PhRvD..83g6011Z,2013arXiv1310.7609H,
2013arXiv1312.7488P,2014arXiv1401.6458B,2014arXiv1403.1987L,
2014arXiv1404.1373A} or a population of millisecond pulsars (MSPs, 
\cite{2011JCAP...03..010A,2013MNRAS.436.2461M}, see also an earlier work 
on a MSP interpretation to EGRET diffuse $\gamma$-ray emission 
\cite{2005MNRAS.358..263W}). Although the DM scenario 
seems very attractive, it is very crucial to investigate the astrophysical 
alternatives of the excess, especially in view that direct detection 
experiments found no signal of DM collision in the corresponding mass 
ranges \cite{2012PhRvL.109r1301A,2014PhRvL.112i1303A}. A first look at the
MSP scenario suggests that it is a plausible interpretation to the data. 
The best-fitting spectrum of the excess is an exponential cutoff power-law,
with power law index $\Gamma\sim1.4-1.6$ and cutoff energy $E_c\sim3-4$ GeV 
\cite{2013PhRvD..88h3521G,2014PhRvD..89f3515M}. All these are consistent 
with the average spectral properties of 
either the Fermi-LAT detected MSPs \cite{2013ApJS..208...17A}, or
globular clusters whose $\gamma$-ray emission is believed to be dominated 
by MSPs \cite{2010A&A...524A..75A}. The number of MSPs 
needed to explain the data is estimated to be a few $\times10^3$ based on 
the observed luminosities of MSPs or globular clusters
\cite{2012PhRvD..86h3511A,2013MNRAS.436.2461M,2013PhRvD..88h3521G}.
Such a number of MSPs is plausible based on the comparison of the
stellar mass content in the Galactic bulge and in the globular 
clusters. The spatial distribution of the $\gamma$-ray excess follows
a gNFW profile, which is somehow expected within the dark matter scenario 
according to N-body simulations with baryon processes 
\cite{2004ApJ...616...16G,2011arXiv1108.5736G}. However, it is interesting
to note that the number distribution of low mass X-ray binaries (LMXBs),
which can be tracers of MSPs, from the central region of Andromeda gives 
a projected $R^{-1.5}$ profile \cite{2007A&A...468...49V,
2007MNRAS.380.1685V}, which is consistent with that to interpret 
the $\gamma$-ray excess \cite{2012PhRvD..86h3511A}.

Hooper et al. investigated in more detail of the MSP scenario to explain the 
GC excess \cite{2013PhRvD..88h3009H}. Based on several assumptions about the 
spatial, spin and luminosity distributions of the MSPs, they claimed that MSPs
cannot explain the $\gamma$-ray excess data without violating the Fermi-LAT
detected number-flux distribution of the MSPs. We revisit this problem
in this work, paying special attention on the assumption of the luminosity
function of MSPs. We will model the spatial and spectral distribution of
MSPs in the Milky Way (MW) disk to reproduce the major MSP observational 
properties as measured by Fermi-LAT, and infer the intrinsic luminosity function
of MSPs (Sec. 2). We then apply the intrinsic luminosity function to a putative
bulge population of MSPs and work out their contribution to the diffuse 
$\gamma$-ray excess without over-producing detectable point sources 
above the sensitivity threshold of Fermi-LAT (Sec. 3). We show that the MSP
scenario can nicely reproduce the $\gamma$-ray excess data, and conclude in 
Sec. 4 with some discussion.

\section{Simulation of MW disk MSPs}

We first try to reproduce the Fermi-LAT observations with a MW disk population 
of MSPs. In the second Fermi-LAT catalog of pulsars (2FPC), 117 pulsars were 
reported, among which 40 are MSPs with 37 having spectral measurements
\cite{2013ApJS..208...17A}. Additionally there are about 30 pulsars 
($\sim$20 are MSPs) which were not included in the 2FPC and can be found
in an online catalog\footnote{https://confluence.slac.stanford.edu/display/GLAMCOG/Public+List+of+LAT-Detected+Gamma-Ray+Pulsars}. Our analysis is based 
on the 37 MSPs in the 2FPC catalog. 

\subsection{Spatial distribution}

The spatial distribution of the MW disk MSPs is adopted as 
\cite{2010JCAP...01..005F}
\begin{equation}
n(r,z)\propto\exp(-r^2/2\sigma_r^2)\exp(-|z|/\sigma_z),
\end{equation}
where $r$ and $z$ are cylindrical coordinates. The radial and vertical
scales are adopted to be the ``base model'' of \cite{2010JCAP...01..005F}, 
with $\sigma_r=5$ kpc and $\sigma_z=1$ kpc. Our study is not very sensitive 
to the spatial distribution, thus we will fix these parameters in the 
following discussion.

\subsection{Spectral distribution}

\begin{figure}[!htb]
\centering
\includegraphics[width=0.48\columnwidth]{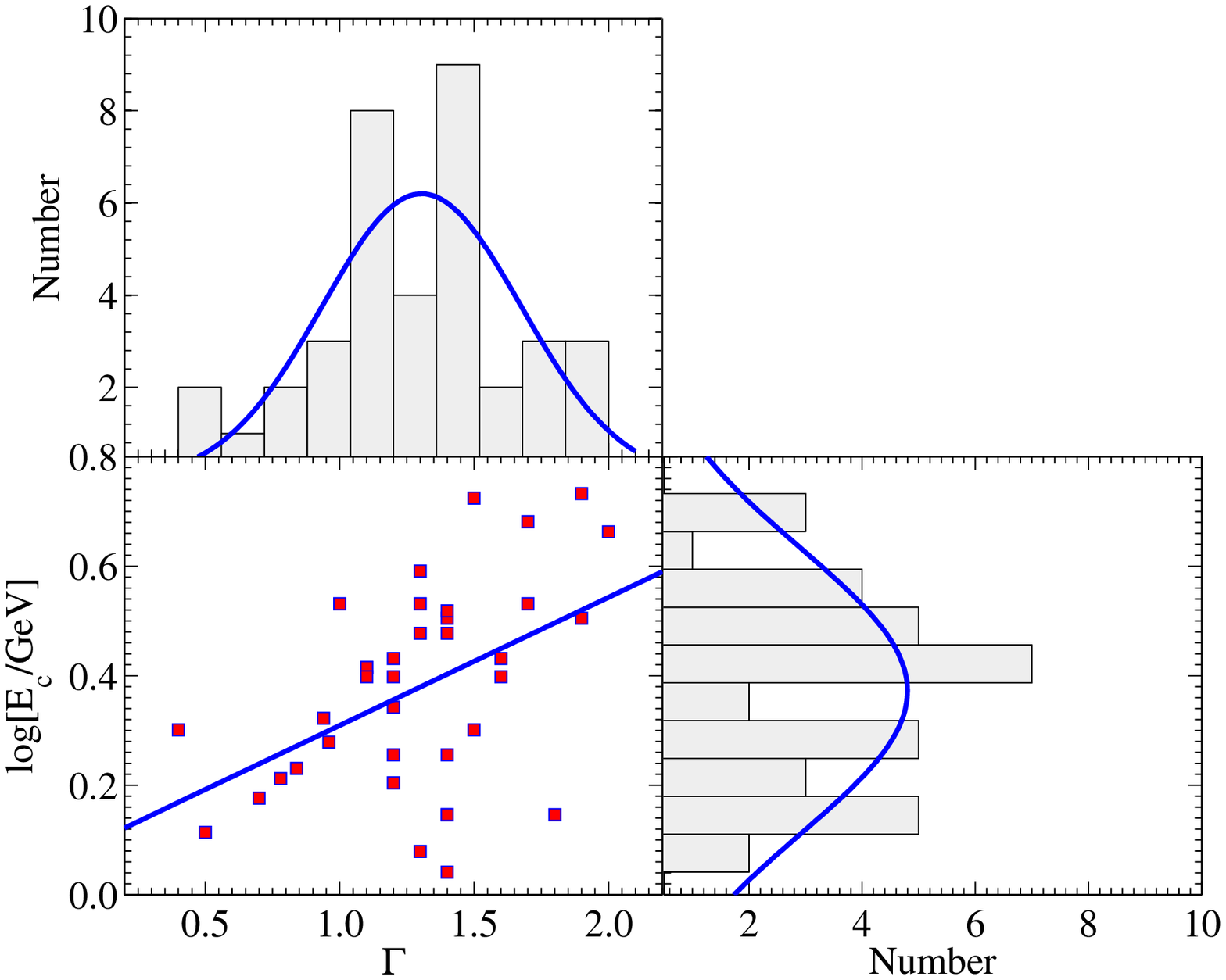}
\includegraphics[width=0.48\columnwidth]{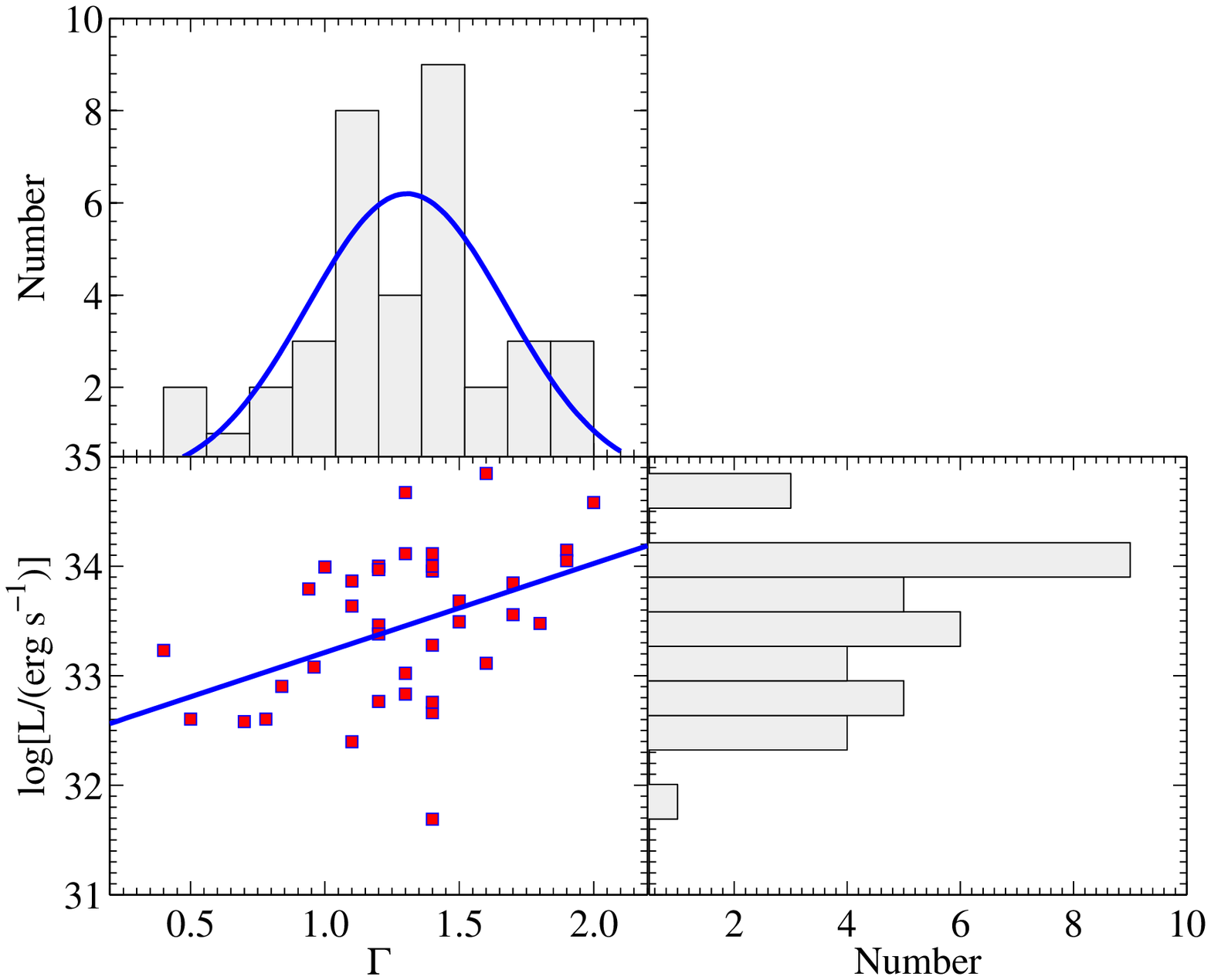}
\includegraphics[width=0.48\columnwidth]{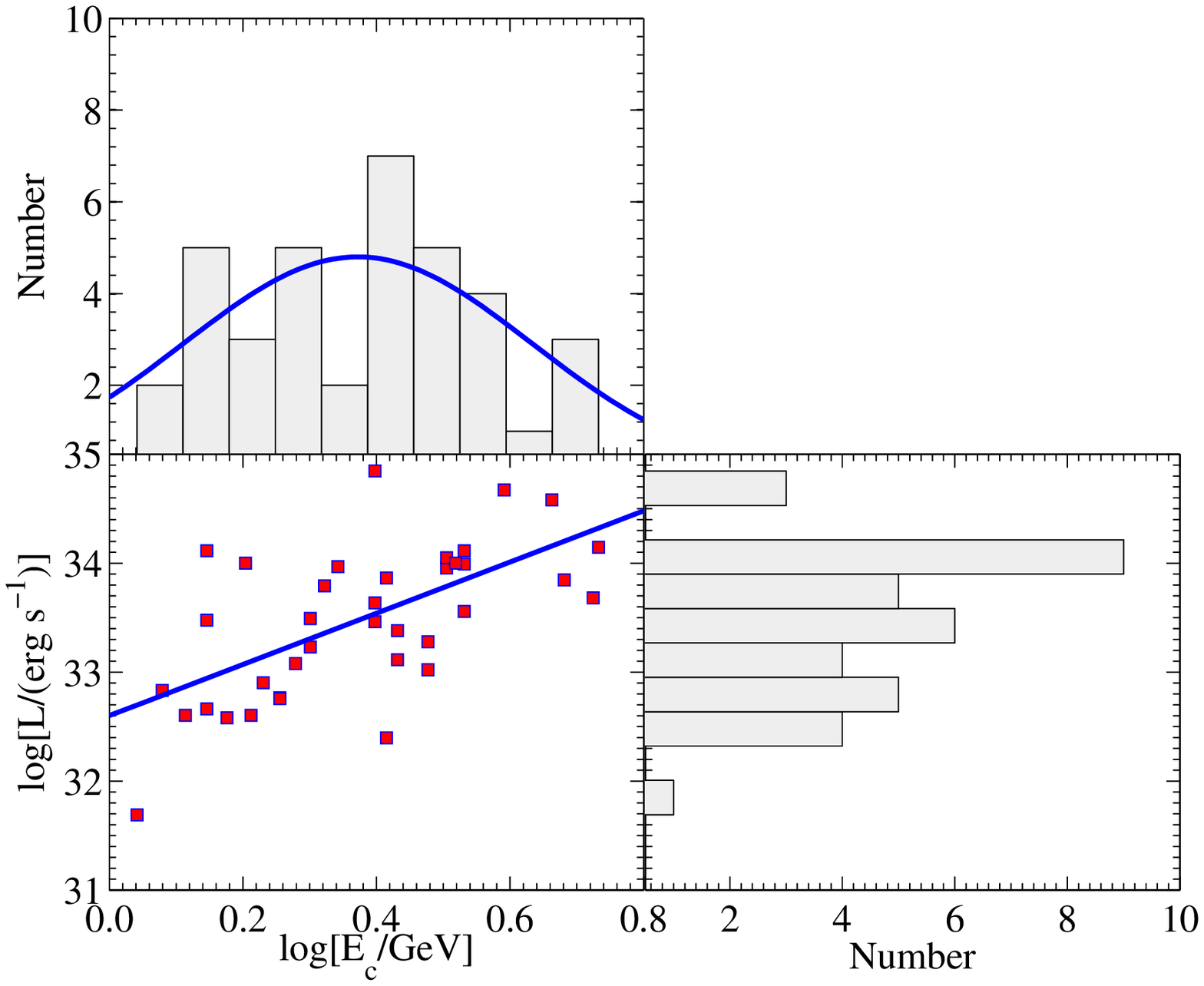}
\caption{Distributions of $\Gamma$, $\log[E_c/{\rm GeV}]$ and
$\log[L/{\rm erg\,s^{-1}}]$ of the Fermi-LAT MSPs.}
\label{fig:FermiMSP}
\end{figure}

The $\gamma$-ray photon spectrum of a MSP can be generally described 
with an exponential cut-off power-law function 
\begin{equation}
{\rm d}N/{\rm d}E\propto E^{-\Gamma}\exp(-E/E_c).
\end{equation} 
Fig. \ref{fig:FermiMSP} shows the distributions of $\gamma$-ray spectral 
indices $\Gamma$, cutoff energies $E_c$ and $\gamma$-ray 
luminosities\footnote{In this work the $\gamma$-ray luminosity and flux 
are computed between 100 MeV and 100 GeV, unless otherwise stated.} $L$ of 
the 2FPC MSPs \cite{2013ApJS..208...17A}. In each panel we have a scatter 
plot to show the correlation between any pair of these parameters, and two 
histograms to show the distributions of each parameter.

The distributions of $\Gamma$ and $\log[E_c/{\rm GeV}]$ can be fitted with 
Gaussian functions (blue lines in the histograms). The mean value and width 
are found to be $1.305$ and $0.370$ for $\Gamma$, and $0.372$ and $0.261$ for 
$\log[E_c/{\rm GeV}]$. We do not fit the luminosity distribution based on
the observed sample because the observational selection effect may favor
the detection of high luminosity ones. The intrinsic luminosity function 
can be only assumed and verified through the observations with a proper
consideration of the detection selection effect. We also note that there 
might be some correlations among these parameters. Linear fittings to 
these correlations give
\begin{eqnarray}
\log[E_c/{\rm GeV}]&=&0.23\Gamma+0.08, \nonumber \\
\log[L/({\rm erg\ s^{-1}})]&=&0.81\Gamma+32.42,\nonumber \\
\log[L/({\rm erg\ s^{-1}})]&=&2.35\log[E_c/{\rm GeV}]+32.58, \nonumber
\end{eqnarray}
but the correlations are weak due to large scatter. The Pearson's 
$r$ values for the three pairs of parameters shown above are 0.47, 0.43 
and 0.62, respectively. For simplicity we will neglect the correlations 
in most of the following discussion. However, the impacts of the 
correlations among these parameters will be tested in the end of Sec. 3. 
In the simulation as discussed below, we will further apply the following 
constraints on the spectral parameters: $\Gamma>0$ and 1 GeV$<E_c<10$ GeV. 

\subsection{Luminosity function}

\begin{figure}[!htb]
\centering
\includegraphics[width=0.7\columnwidth]{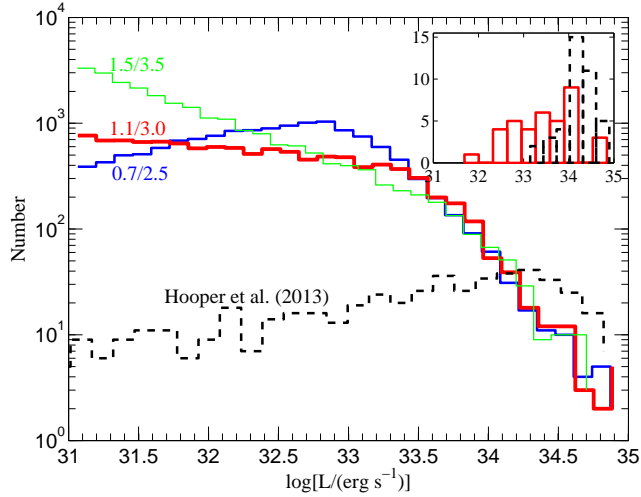}
\caption{Gamma-ray luminosity function (proportional to ${\rm d}N/{\rm d}
\log L$) of MSPs. Solid lines are the broken power-law functions assumed 
in this work for several different sets of parameters as labelled, and the 
dashed line is an example as adopted in \cite{2013PhRvD..88h3009H} with 
$B_0=10^{8.5}{\rm G}$. The total number of the sources of each model is 
normalized to reproduce the observed sample. Inserted is a comparison of 
the luminosity distributions of the Fermi sample (red-solid line) and the 
expectation from the luminosity function given in \cite{2013PhRvD..88h3009H} 
(black-dashed line). See the text for details.}
\label{fig:LF}
\end{figure}

The luminosity function is most relevant for this study. However, it 
cannot be directly derived through the observational sample due to the
sensitivity limit of the detectors. Hooper et al. assumed a power-law 
distribution of the MSP periods\footnote{As shown in Sec.
\ref{sec:LFphysics} below, the ${\rm d}N/{\rm d}P$ dependence would 
significantly affect the shape of luminosity function. This particular
form lacks a physical justification, and cannot account for the observed
$P$ distribution of MSPs.} ${\rm d}N/{\rm d}P\propto P^{-2}$, and a 
constant fraction of the spin-down power goes into $\gamma$-ray 
luminosities $L_{\gamma}\propto \dot{E}$ \cite{2013PhRvD..88h3009H}. 
For a constant magnetic field $B$ one has $\dot{E}\propto P^{-4}$, and the 
luminosity function is ${\rm d}N/{\rm d}L\propto L^{-3/4}$. A log-normal 
distribution of the magnetic field of MSPs is assumed 
\cite{2013PhRvD..88h3009H}, and the resulting luminosity function can be 
derived through a Monte Carlo simulation. An example adopted in 
\cite{2013PhRvD..88h3009H}, with a central value of magnetic field 
$B_0=10^{8.5}{\rm G}$ and a logarithmic standard width $0.2$, is shown by 
the dashed line in Fig. \ref{fig:LF}. We see that such a luminosity function 
is very hard, which might be the reason why Hooper et al. did not find 
enough contribution from low-luminosity MSPs to explain the observed 
$\gamma$-ray excess \cite{2013PhRvD..88h3009H}.

However, we find that such a luminosity function may be over hard. 
If we accept such a luminosity function, and apply the detection
threshold condition\footnote{Flux calculated assuming a unified 
$\gamma$-ray energy spectrum ${\rm d}N/{\rm d}E_{\gamma}\propto 
E_{\gamma}^{-1.46}\exp(-E_{\gamma}/3.3\,{\rm GeV})$.} 
$F(>{\rm GeV})>4\times10^{-10}$ cm$^{-2}$s$^{-1}$,
we find that $\sim 40$ sources could be detected by Fermi-LAT as 
individual MSPs. The luminosities of those $40$ sources are mainly 
above $10^{34}$ erg s$^{-1}$ (dashed histogram in the inset of Fig. 
\ref{fig:LF}), which cannot reproduce the $L$ distribution of the observed 
MSP population (solid, red histogram in the inset of Fig. \ref{fig:LF}).
This suggests that the assumed luminosity function is too hard. We then 
introduce a softer luminosity function. We assume a broken power-law 
form of the luminosity function
\begin{equation}
{\rm d}N/{\rm d}L\propto L^{-\alpha_1}\left[1+(L/L_{\rm br})^2\right]
^{(\alpha_1-\alpha_2)/2}.
\end{equation}
The parameters $\alpha_1$, $\alpha_2$, $L_{\rm br}$ and the normalization
are free parameters, which are determined by reproducing the observed 
sample of MSPs by Fermi-LAT.
To compare with the Fermi-LAT detectability, we apply a latitude dependent 
sensitivity of Fermi-LAT as $F_{\rm th}(>100\,{\rm MeV})=
[2.0\exp(-|b|/10^{\circ})+0.4]\times10^{-8}$ cm$^{-2}$ s$^{-1}$, which
approximately accounts for the effect of the Galactic diffuse background
on the point source sensitivity \cite{2009ApJ...697.1071A}. Here we adopt
a one-year sensitivity of Fermi-LAT, although the 2FPC catalog was based 
on three-year observations. In principle, the sensitivity of Fermi-LAT 
would be better for a pulsar-like spectrum, which is harder than the 
$E^{-2}$ spectrum used to derive the above point source sensitivity  
\cite{2013ApJS..208...17A}. On the other hand, identifying a MSP would
be challenging if the flux is just above the sensitivity threshold,
since enough photons are needed to conduct MSP timing studies.
The flux limit of identified MSPs is somewhat higher than the
point source detection sensitivity, and we adopt a more conservative 
detection threshold to mimic the threshold for identifying a MSP.
This adopted detection threshold is also close to the upper edge of 
the three-year sensitivity bands for point sources with pulsar-like 
spectrum given in Fig. 17 of \cite{2013ApJS..208...17A}.

\subsection{Results}

\begin{figure}[!htb]
\centering
\includegraphics[width=0.48\columnwidth]{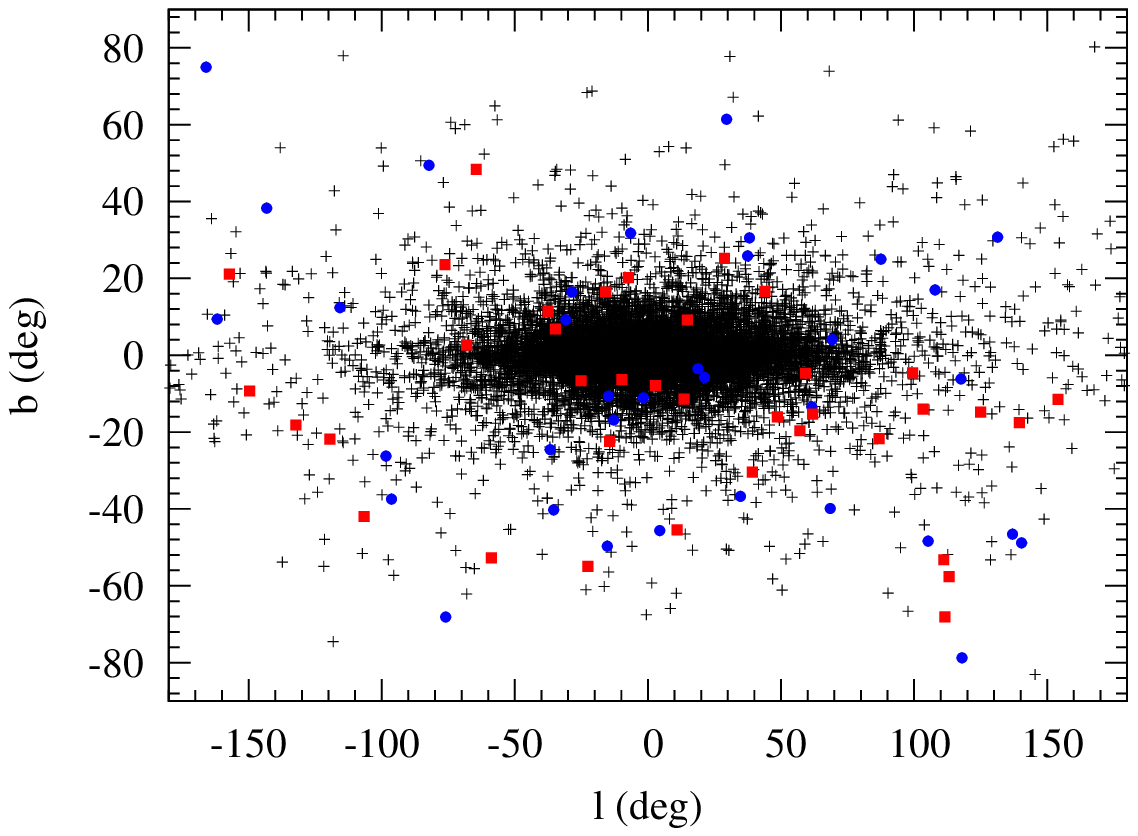}
\includegraphics[width=0.48\columnwidth]{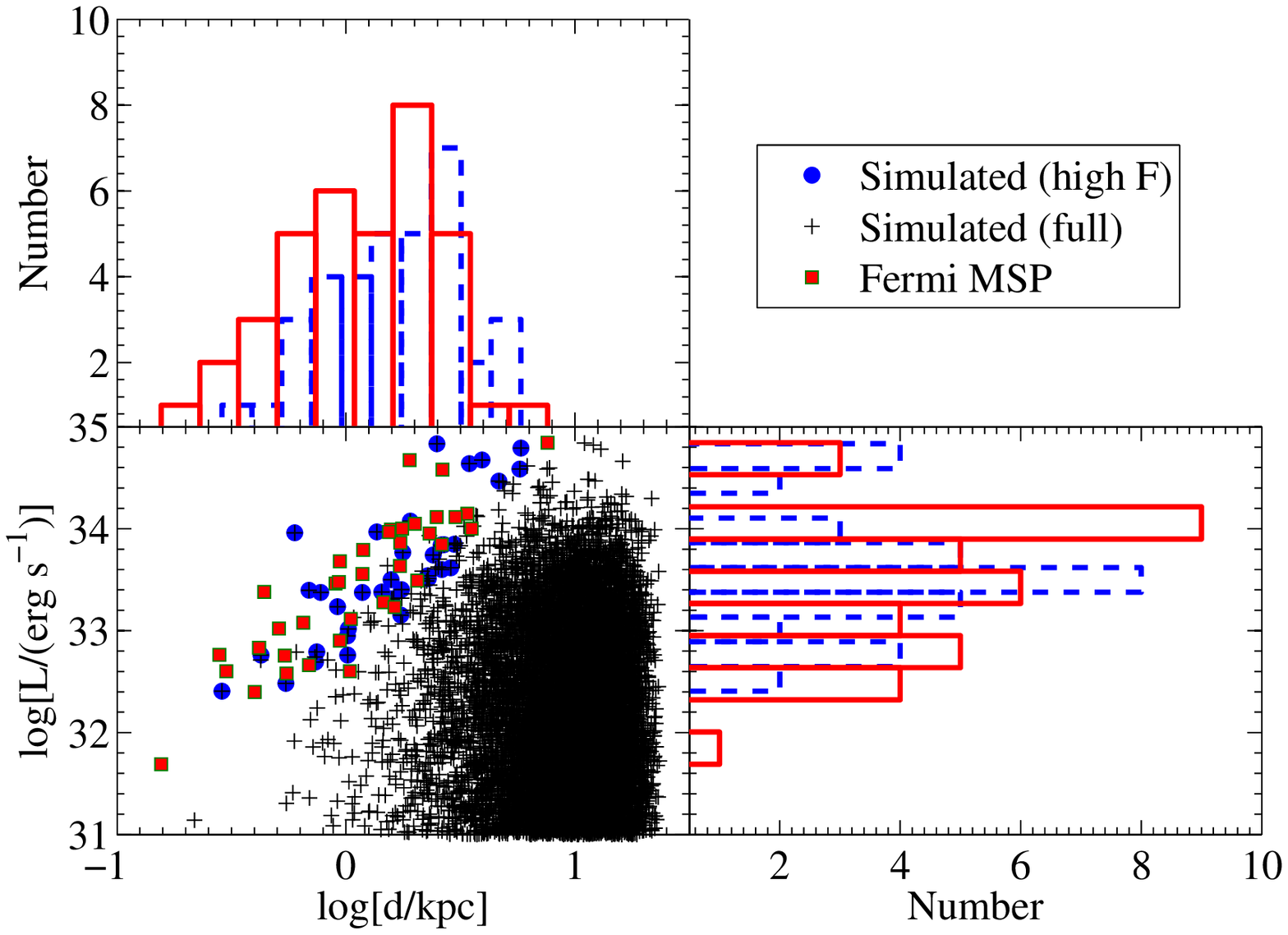}
\includegraphics[width=0.48\columnwidth]{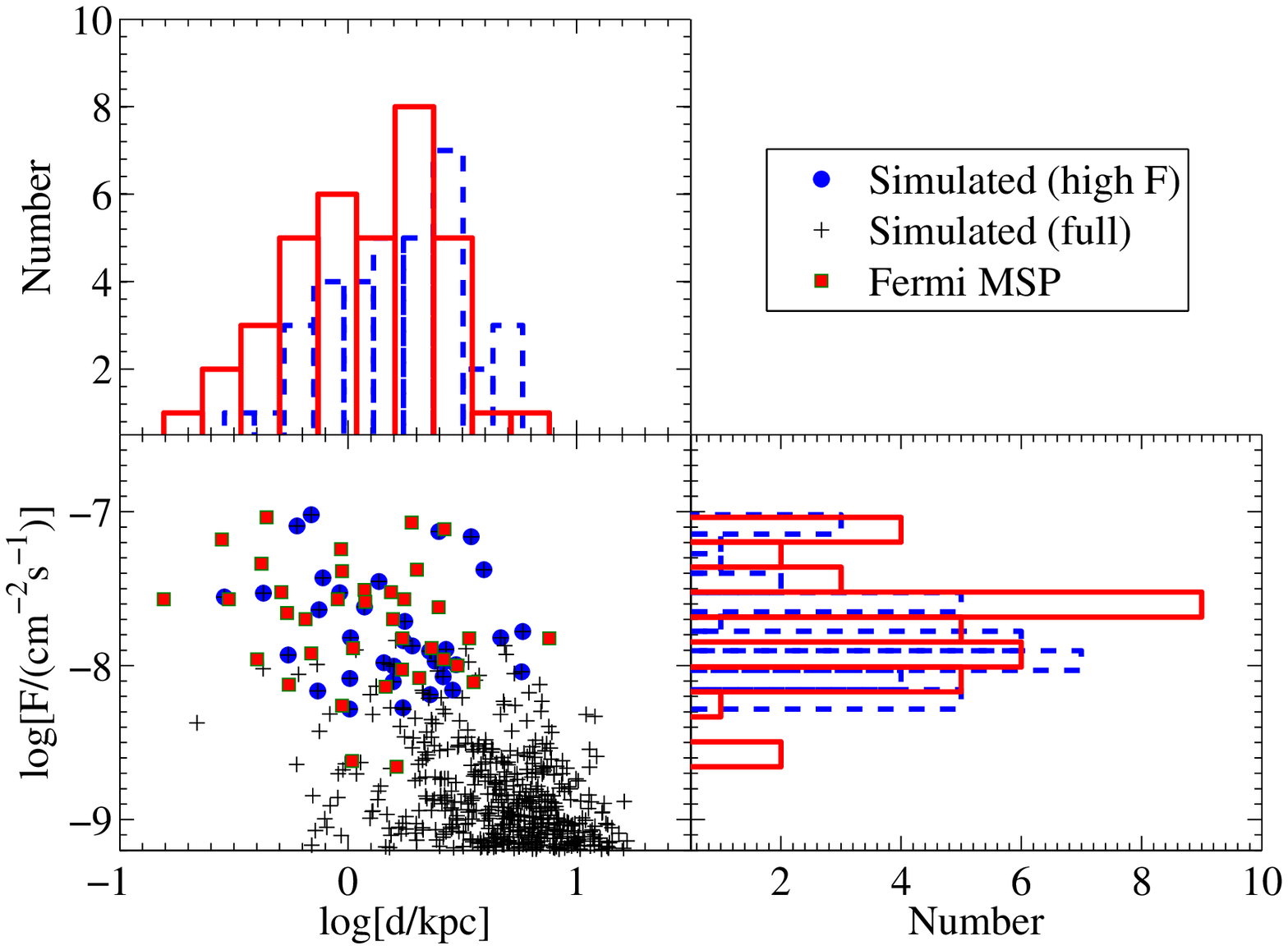}
\includegraphics[width=0.48\columnwidth]{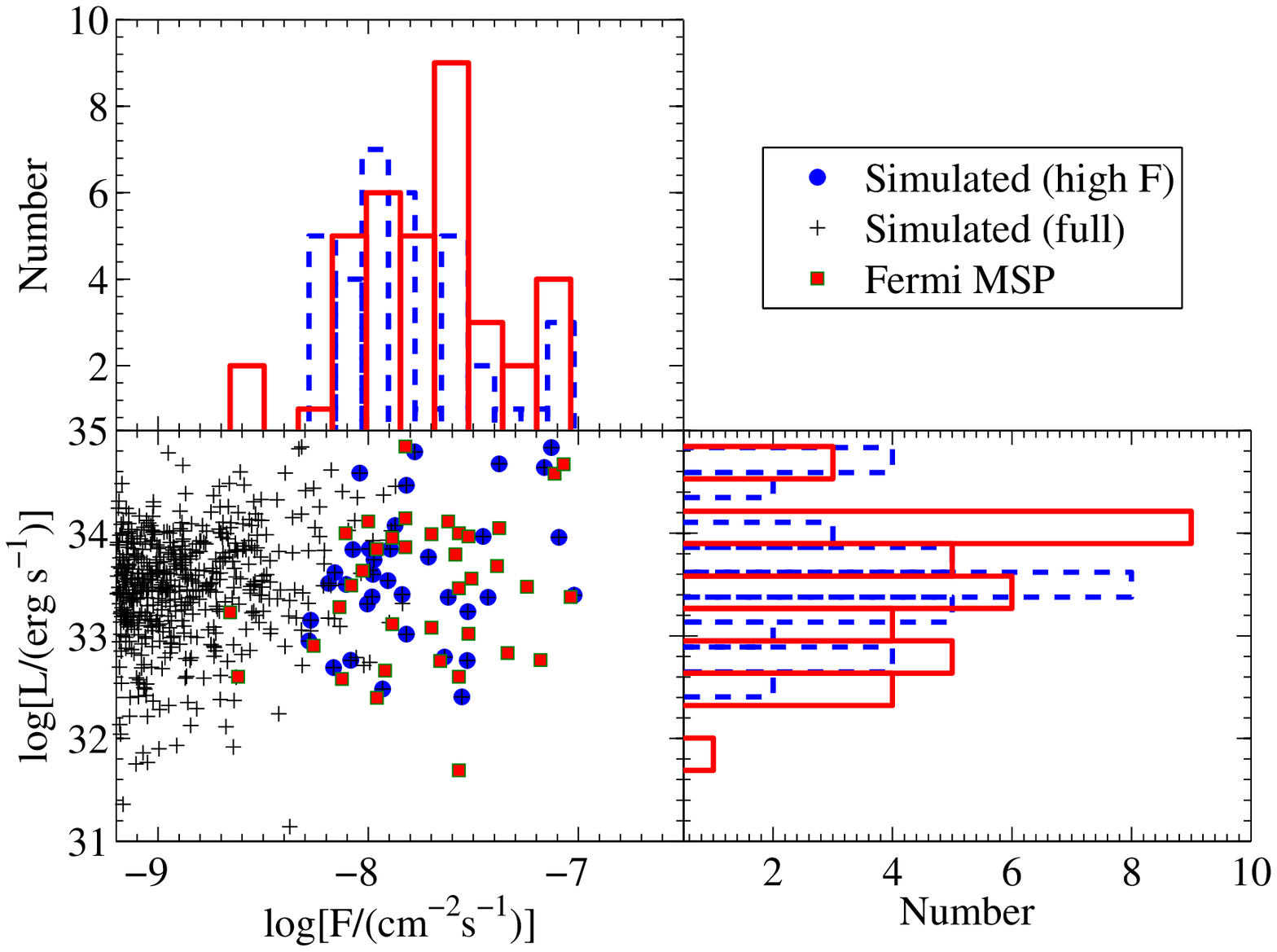}
\caption{Comparisons of the spatial distribution, distances, luminosities 
and fluxes of MSPs between the simulated sample and the Fermi-LAT detected 
sample. The black crosses are the full simulated sample, the blue dots
are the high-flux ones with $F>F_{\rm th}$, and the red squares 
are Fermi-LAT detected sample. Histograms in the last three panels show
the distributions of related quantities for the simulated high flux sample
(blue dashed) and the Fermi-LAT sample (red solid).}
\label{fig:simu_MW}
\end{figure}

With the above mentioned spatial distribution, spectral distribution
and luminosity function, we can simulate MSPs in the MW disk. The number 
of the simulated sources is normalized to reproduce the detected number 
of MSPs with fluxes (above 100 MeV) larger than $F_{\rm th}(b)$. The results 
from one realization with luminosity function parameters $\alpha_1/\alpha_2=
1.1/3.0$ and $L_{\rm br}=4\times10^{33}$ erg s$^{-1}$ are shown in
Fig. \ref{fig:simu_MW}. The top-left panel shows the sky distribution,
and other panels show the distributions of distance $d$, luminosity $L$ 
and flux $F$, respectively. In each panel, the black crosses represent 
the full simulated sample, the blue dots are the simulated sample with 
fluxes above $F_{\rm th}$, and the red squares are Fermi-LAT detected sample. 
The distributions of $\log d$, $\log L$ and $\log F$ are shown by the
histograms in the rest three panels for the simulated high flux sample 
and the Fermi-LAT sample. We can see from this figure that the model can 
roughly reproduce the Fermi-LAT observations. To be more quantitative,
we check the consistency between the simulated sample and the observed
sample using the Kolmogorov-Smirnov test method. The probabilities that
these two samples come from the same distributions are about $0.49$, 
$0.93$ and $0.44$ for the distance, luminosity and flux distributions, 
respectively. The total number of the MW MSPs in this simulation is 
$\sim6000$ for $L>10^{32}$ erg s$^{-1}$, which gives $\sim40$ detectable 
sources.

Due to the limited statistics of the observed sample, the model parameters
cannot be precisely determined. We have tested other parameters and 
find that changing the luminosity function parameters $\alpha_1/\alpha_2$ 
from $0.7/2.5$ to $1.5/3.5$ (accordingly $L_{\rm br}$ changes from 
$1.0\times10^{33}$ to $1.0\times10^{34}$ erg s$^{-1}$) would not 
significantly affect the model results. The observational distributions 
can all be approximately reproduced, even though the overall 
agreement becomes worse than the best fit results. 
The parameters of some luminosity function models are listed in Table 
\ref{table:para}, and the three example luminosity functions 
are plotted in Fig. \ref{fig:LF}. Compared with the luminosity function 
adopted in \cite{2013PhRvD..88h3009H}, our derived luminosity functions 
give many more low-luminosity sources, which would contribute significantly
to the unresolved diffuse emission.

\begin{table}[!htb]
\centering
\caption{Parameters adopted in the simulation.}
\vspace{2mm}
\begin{tabular}{cccc}
\hline \hline
$\alpha_1/\alpha_2$ & $L_{\rm br}$ & $N_{\rm MW}^{a}$ & $N_{\rm bulge}^{a}$ \\
 & ($10^{33}$ erg s$^{-1}$) & & \\
\hline
$0.7/2.5$ & $1.0$ & $9000$ & $17000$ \\
$1.1/3.0$ & $4.0$ & $6000$ & $13000$ \\
$1.5/3.5$ & $10.0$ & $8000$ & $16000$ \\
\hline
\hline
\end{tabular}\\
$^a$Number with $L$ between $10^{32}$ and $10^{35}$ erg s$^{-1}$.
\label{table:para}
\end{table}

\subsection{Physical interpretation of the luminosity function}
\label{sec:LFphysics}

We have shown that with a broken power-law form of the intrinsic 
luminosity function Eq. (3) and a proper consideration of the detection 
threshold, the observational properties of the MW disk MSPs can be well 
reproduced. It is, however, necessary to justify that such a luminosity 
function is reasonable in realistic MSP models. Many theoretical models 
have been proposed to interpret the $\gamma$-ray emission from pulsars, 
including the polar cap model \cite{1996ApJ...458..278D}, outer 
gap model \cite{1986ApJ...300..500C}, slot gap model
\cite{2004ApJ...606.1143M}, 
pair starved polar cap model \cite{2004ApJ...617..471M},  
two-pole caustic model \cite{2003ApJ...598.1201D}, and
annular ring model \cite{2004ApJ...606L..49Q}. A recent study of the
$\gamma$-ray light curves of MSPs seems to favor the outer gap model
or two-pole caustic model, although other models may also work for
some cases \cite{2014arXiv1404.2264J}. In this work we do not
get into the detailed emission models of MSPs. Rather, we perform
a pheonomenological model to study the statistical properties of 
the $\gamma$-ray MSPs. As shown below, the intrinsic luminosity
function does not sensitively depend on detailed emission models.

One can write the intrinsic luminosity function in the form
\begin{equation}
\frac{{\rm d}N}{{\rm d}L} = \frac{{\rm d}N}{{\rm d}P} \cdot 
\frac{{\rm d}P}{{\rm d} \dot E} \cdot \frac{{\rm d} \dot E}{{\rm d}L}.
\end{equation}
We can see that it depends on the period distribution ${\rm d}N/{\rm d}P$, 
period-dependent spin-down luminosity ${\rm d}P/{\rm d} \dot E$, and the 
fraction of spin-down luminosity that goes to the observed $\gamma$-ray 
luminosity ${\rm d} \dot E/{\rm d}L$. Taking roughly a constant magnetic 
field strength for MSPs (so that $\dot E \propto P^{-4}$), and assuming 
$L\propto\dot{E}^a$ and $dN/dP\propto P^b$ ($b=-2$ as adopted in
\cite{2013PhRvD..88h3009H}), it is straightforward to derive
\begin{equation}
\frac{{\rm d}N}{{\rm d}L}\propto L^{-\frac{(b+1)}{4a}-1}.
\end{equation}
Therefore the indices $a$ and $b$ determines the power law
index of the luminosity function. The slope of luminosity
function is much more sensitive to $b$ than to $a$.

The $L-\dot{E}$ relation depends on pulsar emission models, see e.g. 
\cite{2002ApJ...576..366H} for polar cap models and 
\cite{2013ApJ...766...98H} for outer gap models. In general,
$a$ is in the range of $0.5-1$. If the index $b$ is a constant,
one cannot reproduce the required broken power-law luminosity
function for typical values of $a$ (due to the insensitivity
of the results on $a$).

The results are more sensitive to $b$. If all the MSPs reach the 
``spin-up'' limit and then spin-down, we may expect $b=1$ if the birth 
rate $\dot{N}$ is constant\footnote{For $\dot{N}=C$, $N\propto\tau\propto 
P/\dot{P}\propto P^2$ if $B$ is constant.}. However, observationally 
we do not see such a behavior. Neither do we see the $b=-2$ behavior 
introduced by \cite{2013PhRvD..88h3009H}. Rather, observationally, the 
period distribution of MSPs is not a single power-law. It has a peak 
around $\sim 3-4$ ms. The deficiency of MSPs with even shorter periods 
is not due to a seletion effect, since they have an even larger $\dot E$ 
and should be more easily detected if they do exist. Therefore the
break in the ${\rm d}N/{\rm d}P$ distribution is intrinsic, and it 
naturally introduces a break in the intrinsic luminosity function of 
MSPs. Physically, there is a maximum spin frequency at birth for MSPs, 
defined by the so-called ``spin-up'' line, at which the accretion from 
the companian can no longer transfer angular momemtum to the pulsar (the 
shortest period of pulsars to date is 1.4 ms \cite{2006Sci...311.1901H}). 
Introducing a distribution of magnetic field strength and a distribution 
of the ``ending time'' during the spin-up phase for MSPs at birth would 
naturally give rise to a peak in the $P$ distribution. We note that for 
a typical value $B\sim 10^{8.5}$ G, such a peak period corresponds to a 
spin-down power $(1.5-5)\times10^{34}$ erg s$^{-1}$. If the $\gamma$-ray 
luminosity of MSPs shares a few percent of $\dot{E}$, it would correspond 
to a break of the luminosity function at $10^{33}$ erg s$^{-1}$, which is 
the one required in our modeling.

The slopes $\alpha_1$ and $\alpha_2$ can be determined by the parameters
$a$ and $b$. We adopt power-law fits to approximate the period distribution
below and above the peak period\footnote{The distribution may also be
fitted as a Gaussian distribution (e.g., \cite{2005MNRAS.358..263W}).} 
$P_{\rm br}$. We have $b\approx-2$ for $P>P_{\rm br}$, and $b\approx 2-3$ 
for $P<P_{\rm br}$. In the low-luminosity regime ($P > P_{\rm br}$),
usually $a \approx 1$. This gives $\alpha_1 \sim 0.75$. In the
high-luminosity regime ($P < P_{\rm br}$), one has $a \approx 0.5-1$ 
\cite{2002ApJ...576..366H,2013ApJ...766...98H}. This gives 
$\alpha_2\approx 1.75-3$. These values of $\alpha_1$ and $\alpha_2$ 
are close to those adopted in Sec. 2.4 in order to reproduce the
Fermi-LAT observations.

\section{Simulation of bulge MSPs}

The Galactic bulge is rich in stars, hence also rich in remnants of stars, 
i.e., compact objects such as black holes and neutron stars.
The number of compact objects is estimated to be $\sim20000$ in the
inner pc region \cite{2000ApJ...545..847M,2007MNRAS.377..897D}.
The number of compact objects should be much more within the kpc scale,
which is relevant to this study. Furthermore, the high number density of 
stars in the Galactic bulge facilitates the dynamic formation 
of binary systems \cite{2007A&A...468...49V,2007MNRAS.380.1685V}, which 
are progenitors of MSPs. In this section we model the MSP population in 
the Galactic bulge, based on the spectral parameters and luminosity function 
derived in Sec. 2. We will investigate their contribution to the GC 
$\gamma$-ray excess without over-predicting detectable sources by Fermi-LAT.

\subsection{Spatial distribution}

MSPs are believed to be recycled pulsars born in binary systems. The LMXBs
are considered as progenitor of MSP systems, and are believed to
trace the distribution of MSPs \cite{1996ASPC..105..547B}. 
The observational surface 
density profile of resolved LMXBs in the center of M 31 (at sub-kpc scale) 
traces the stellar mass profile at a scale larger than $1'$, and shows a 
significant ``excess'' in the inner arc-minute region 
\cite{2007A&A...468...49V}. The inner ``excess'' can be explained by the 
dynamic formation scenario of LMXBs through stellar encounters in a very 
high stellar density environment \cite{2007MNRAS.380.1685V}. 
The overall surface density profile of LMXBs can be approximated with 
a $\theta^{-1.5}$ behavior, which is consistent with the projected 
profile of the $\gamma$-ray excess in the GC region 
\cite{2012PhRvD..86h3511A}.

\begin{figure}[!htb]
\centering
\includegraphics[width=0.48\columnwidth]{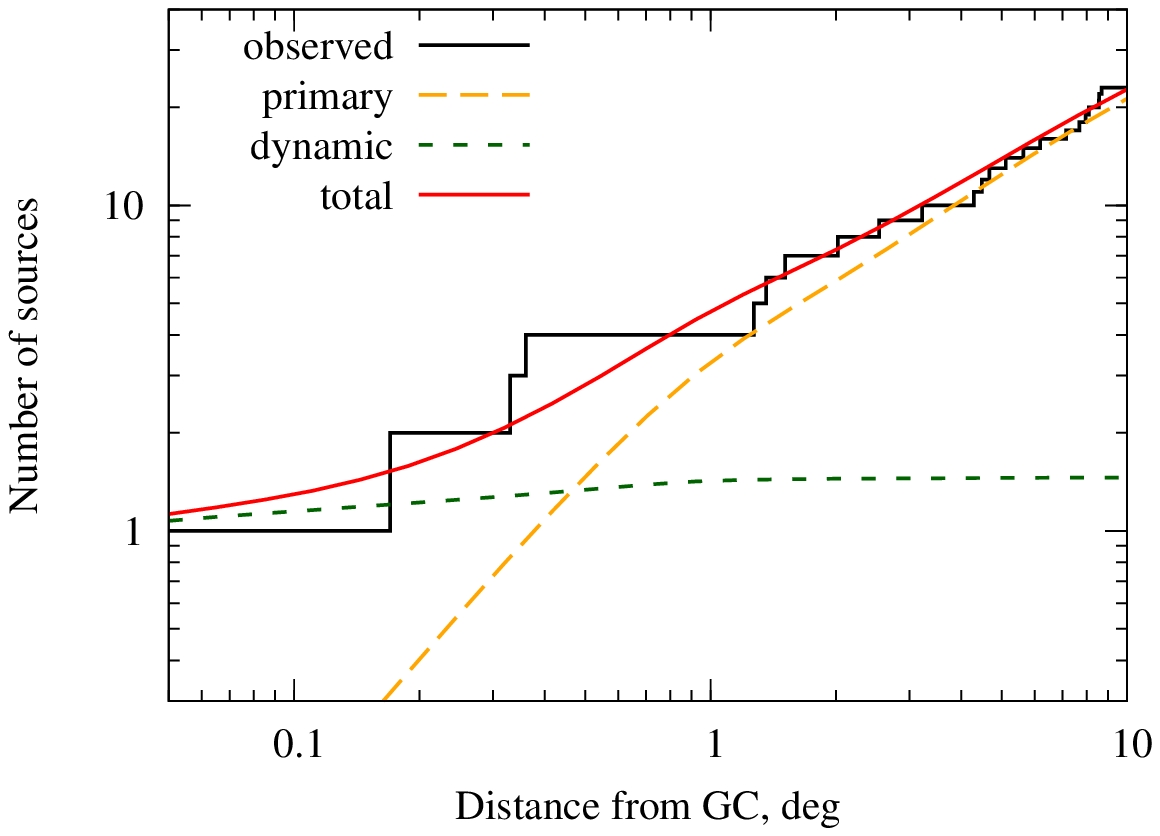}
\includegraphics[width=0.48\columnwidth]{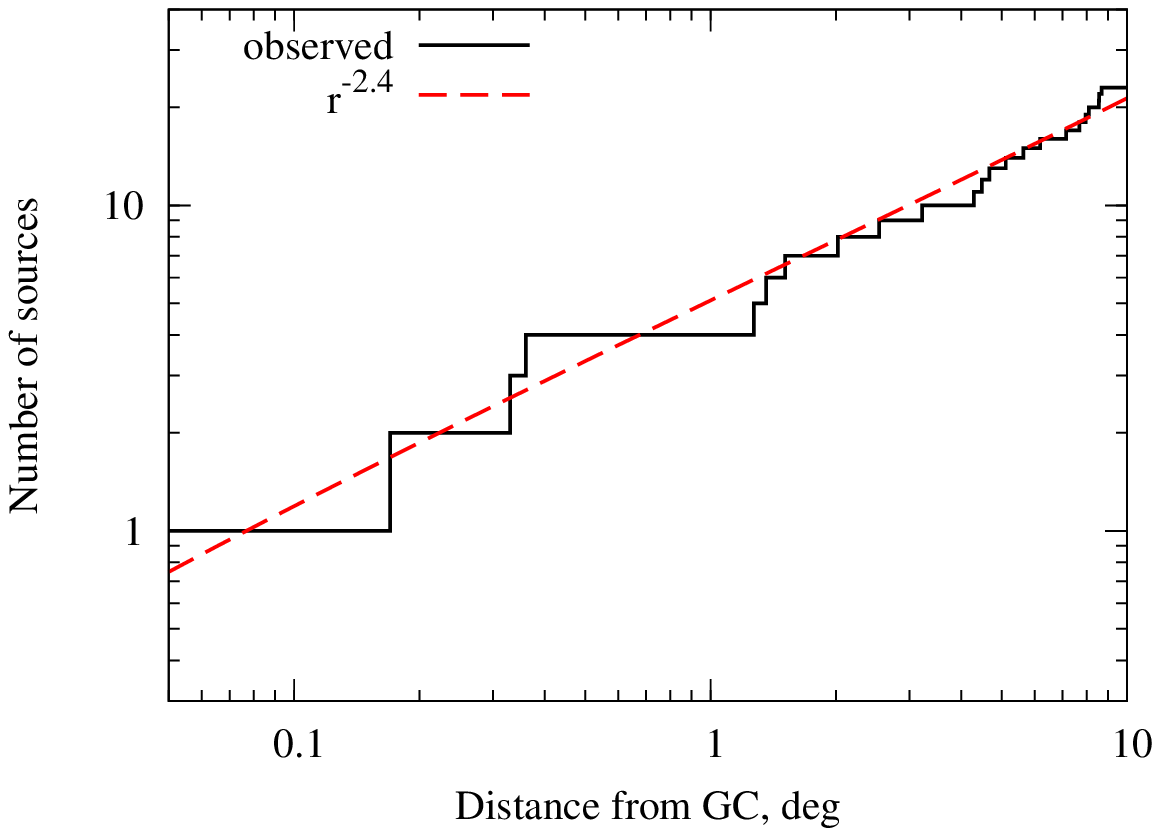}
\caption{Left: spatial distribution of the MW transient LMXBs 
\cite{2008A&A...491..209R}, compared with the prediction from the stellar 
mass distribution. The primary term is proportional to $\rho_{\star}$ and 
the dynamic term is proportional to $\rho_{\star}^2$ 
\cite{2007A&A...468...49V,2007MNRAS.380.1685V}. Right: spatial distribution 
of the MW LMXBs compared with a $r^{-2.4}$ distribution which mimics the
generalized-NFW square profile with inner slope $\gamma=1.2$.}
\label{fig:lmxb}
\end{figure}

The distribution of MW LXMBs is less well constrained. Revnivtsev et al. 
reported a number of LMXBs within the central 10 degrees of the MW
\cite{2008A&A...491..209R}. The cumulative number of transient LMXBs also 
shows an increase in the innermost region compared with the stellar mass
distribution $\rho_{\star}$. We show that the model prediction can well
reproduce the data by adding a dynamic term which is proportional to 
$\rho^2_{\star}$, as shown in the left panel of Fig. \ref{fig:lmxb}.
The stellar mass model we use is the same as that used in
\cite{2008A&A...491..209R}. In the right panel of Fig. \ref{fig:lmxb}
we show the expected cumulative distribution of LMXBs for a $r^{-2.4}$ 
profile as indicated by the GeV $\gamma$-ray excess, where $r$ is the 
spherical coordinate. It is intriguing to see that the observed 
spatial distribution of LMXBs can be also nicely fitted by the $r^{-2.4}$ 
generalized NFW profile. This suggests that an apparent generalized
NFW profile does not necessarily mean a dark matter signature.

\begin{figure}[!htb]
\centering
\includegraphics[width=0.7\columnwidth]{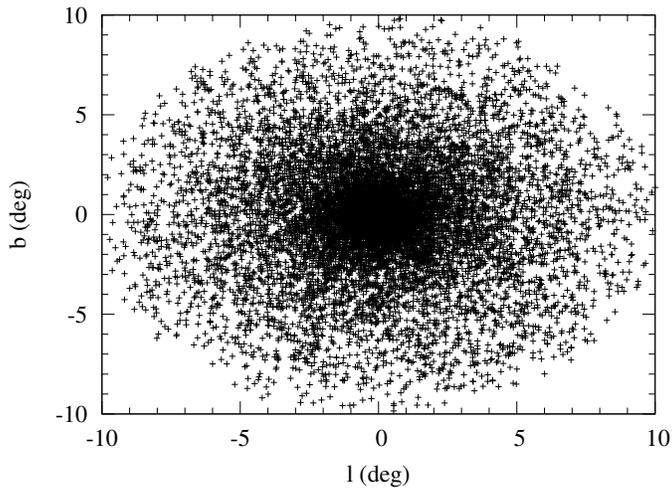}
\caption{Simulated spatial distribution of the bulge MSPs.}
\label{fig:simu_bulge_sky}
\end{figure}

For simplicity in the following we  
assume a spherically symmetric distribution with
spatial profile $r^{-2.4}$ of MSPs in the bulge. Note that if the
stellar model of LMXB formation is correct, there should be an
asymmetry in the MSP distribution, with a tendency to elongate along the 
Galactic longitude. According to the observed LMXB sample, there is no such 
elongation at least for the the central $1.3^{\circ}$ away from the GC
\cite{2008A&A...491..209R}. For a better determination of the density
profile of MSPs, we need a larger sample of LMXBs. On the other hand, 
the search for asymmetry of the GC $\gamma$-ray excess cannot exclude 
a weak elongation along the Galactic plane, although no significant 
asymmetry was found \cite{2014arXiv1402.6703D}. This means that even if 
the MSP distribution is slightly asymmetric, it may also be consistent with 
the morphology of the observed $\gamma$-ray excess. The simulated spatial 
distribution of the bulge MSPs is shown in Fig. \ref{fig:simu_bulge_sky}.
Note here we additionally apply a truncation of the density profile for
$\theta>10^{\circ}$, which may represent the size of the bulge.

\subsection{Spectral distribution}

\begin{figure}[!htb]
\centering
\includegraphics[width=0.7\columnwidth]{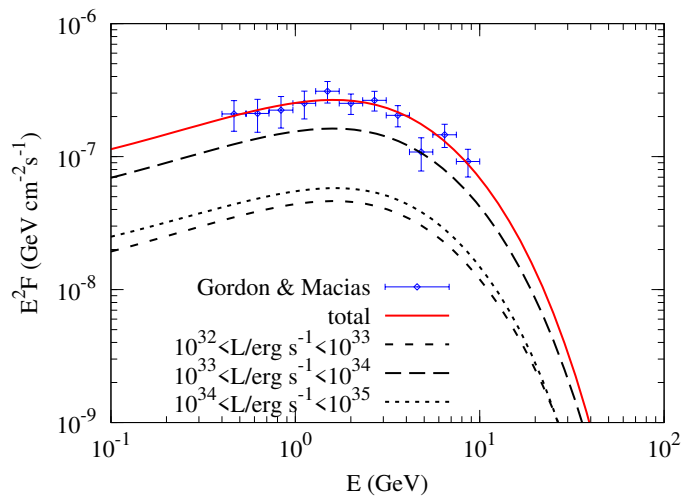}
\caption{Cumulative spectrum of the bulge MSPs compared with the Fermi-LAT
GC excess data \cite{2013PhRvD..88h3521G}.}
\label{fig:simu_bulge_spec}
\end{figure}

Finally we compare the expected $\gamma$-ray spectrum from the population
of bulge MSPs with the Fermi-LAT GC excess. The cumulative flux from the MSP
population depends on the number of MSPs, which is adjusted to match the 
$\gamma$-ray excess data \cite{2013PhRvD..88h3521G}. 
Fig. \ref{fig:simu_bulge_spec} shows the result for the model with the 
luminosity function parameters $\alpha_1/\alpha_2=1.1/3.0$ as given in Table 
\ref{table:para}. To compare with the data, only the MSPs that lie within the
$7^{\circ}\times 7^{\circ}$ box centered on GC are employed. It is not 
surprising that the model can well reproduce the data, because the average 
energy spectra of MSPs are consistent with the $\gamma$-ray excess data. 
It also shows that for this luminosity function
the MSPs with luminosities between $10^{33}$ and $10^{34}$ erg s$^{-1}$
contribute dominantly to the total flux. This is reasonable because the
break of the luminosity function lies in this luminosity range. 
The number of MSPs needed to give enough cumulative flux to explain the 
data is estimated to be $\sim13000$ for $L>10^{32}$ erg s$^{-1}$. Obviously 
such a number depends on the lower cutoff of the luminosity function.
For the other two luminosity functions in Table \ref{table:para} we have
similar results, with quantitatively different number of sources and 
weights among different luminosity ranges.

\begin{figure}[!htb]
\centering
\includegraphics[width=0.7\columnwidth]{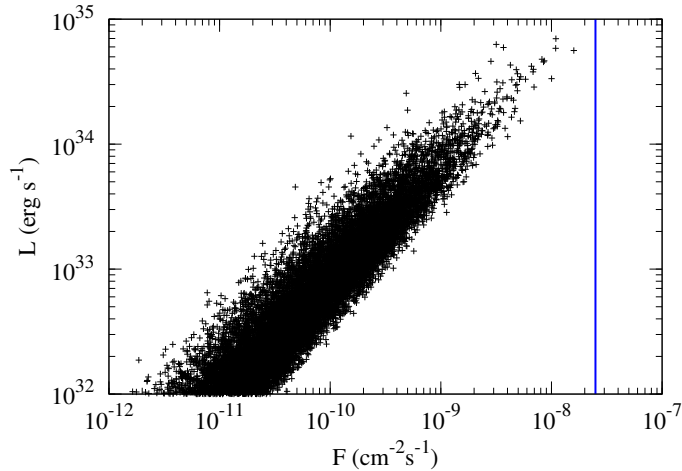}
\caption{Simulated $F$ vs. $L$ distributions of the bulge MSPs. The 
vertical line is the sensitivity of Fermi-LAT for sources located in the 
Galactic plane \cite{2009ApJ...697.1071A}.}
\label{fig:simu_bulge_FL}
\end{figure}

In order to check whether the bulge MSP population violates the Fermi-LAT
observations, we show the fluxes versus luminosities of these MSPs 
in Fig. \ref{fig:simu_bulge_FL}. The vertical line is the sensitivity of 
Fermi-LAT for sources located in the Galactic plane 
\cite{2009ApJ...697.1071A}. It is shown that none of these bulge MSPs 
could be detected as an individual source by Fermi-LAT, which means that
all of them should contribute to the diffuse emission.

We can compare the number of MSPs estimated here with that derived in other 
works. Using the average luminosity of the Fermi-LAT detected MSPs, $\bar{L}
\approx10^{34}$ erg s$^{-1}$, Macias \& Gordon estimated a number of 
$\sim1000$ MSPs in order to explain the data \cite{2014PhRvD..89f3515M}. 
This number should be a lower bound because there should be more 
low-luminosity MSPs which are not detected. In our work, the main
contribution to the total $\gamma$-rays comes from the MSPs with 
luminosities between $10^{33}$ and $10^{34}$ erg s$^{-1}$ (Fig. 
\ref{fig:simu_bulge_spec}). We find that the number of MSPs in this
luminosity range is about $4200$ for a total number of $13000$ 
($L>10^{32}$ erg s$^{-1}$). If we count only the $7^{\circ}\times
7^{\circ}$ box the number becomes $2700$, which is consistent with
the lower limit derived in \cite{2014PhRvD..89f3515M}, given the
average luminosity is about several times smaller. However, as we have
mentioned, this number depends on how many low-luminosity MSPs there 
are. There is only one MSP with luminosity below $10^{32}$ erg s$^{-1}$
in the Fermi-LAT sample, but we are not sure whether the luminosity function 
can extend to even lower luminosities or not. If so the number of MSPs
may be even larger than that given in Table \ref{table:para}.

\begin{figure}[!htb]
\centering
\includegraphics[width=0.7\columnwidth]{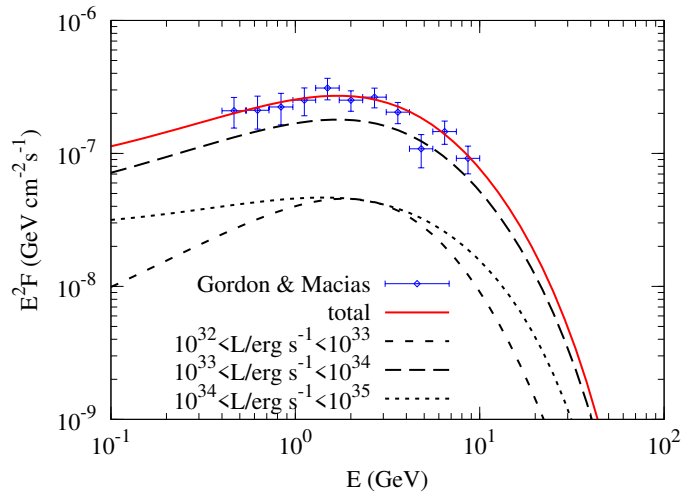}
\caption{Same as Fig. \ref{fig:simu_bulge_spec} but the correlations
between $L$ and $\Gamma$, $L$ and $E_c$ are included. See the text for
details.}
\label{fig:simu_bulge_spec2}
\end{figure}

Finally we test the model with the correlations between spectral parameters 
and luminosity. We employ a simple approach to approximate the correlations 
between $L$ and the spectral parameters shown in Fig. \ref{fig:FermiMSP}: 
for $10^{32}<L<10^{33}$ erg s$^{-1}$, 
$\langle\Gamma\rangle=1.0$, $\langle\log[E_c/{\rm GeV}]\rangle=0.2$, 
for $10^{33}<L<10^{34}$ erg s$^{-1}$, $\langle\Gamma\rangle=1.3$, 
$\langle\log[E_c/{\rm GeV}]\rangle=0.4$, and for $10^{34}<L<10^{35}$ 
erg s$^{-1}$, $\langle\Gamma\rangle=1.6$, 
$\langle\log[E_c/{\rm GeV}]\rangle=0.6$, respectively. Here the angle 
brackets $\langle...\rangle$ denote the average valus of corresponding
quantities. The widths of $\Gamma$ and $\log[E_c/{\rm GeV}]$ are kept 
unchanged. The result for the same model as that in Fig. 
\ref{fig:simu_bulge_spec} is shown in Fig. \ref{fig:simu_bulge_spec2}.
Since the MSPs with $10^{33}<L<10^{34}$ erg s$^{-1}$ dominate the
contribution, the total spectrum do not change significantly compared
with that when the correlations are not taken into account.

\section{Conclusion and discussion}

The analysis of the Fermi-LAT data revealed symmetric and extended 
$\gamma$-ray excess in the GC region peaking at GeV energies
\cite{2009arXiv0910.2998G,2009arXiv0912.3828V,2011PhLB..697..412H,
2011PhLB..705..165B}. The origin of the excess is not clear, and the 
promising scenarios include DM annihilation and an unresolved MSP 
population. Although the spectrum of the $\gamma$-ray excess is quite
consistent with the average spectrum of the Fermi-LAT detected MSPs,
it was argued that in order not to over-produce the detectable MSPs 
by Fermi-LAT, the unresolved MSP population can only account for
$\lesssim10\%$ of the observed $\gamma$-ray \cite{2013PhRvD..88h3009H}.

In this work we study the MSP scenario in detail, by including more 
comprehensive observational constraints from the observational properties 
of the Fermi-LAT detected MSP sample. We find that there is a large 
uncertainty in the intrinsic $\gamma$-ray luminosity function of MSPs, 
which affects significantly the prediction of the diffuse emission from 
the unresolved MSP population. It was found that the luminosity function 
adopted in \cite{2013PhRvD..88h3009H} might be too hard to reproduce the 
observed luminosity function of the Fermi-LAT MSP sample. Adjusting properly 
the intrinsic luminosity function we can well reproduce the observational 
properties of the Fermi-LAT MSPs with the MW population of MSPs. Based on 
this refined luminosity function, we find that a population of MSPs in the
bulge can be enough to explain the $\gamma$-ray excess without over-producing 
the detectable MSPs above the sensitivity of Fermi-LAT.
The number of MSPs with luminosities higher than $10^{32}$ erg s$^{-1}$
in the whole bulge region is estimated to be $(1-2) \times 10^{4}$ in order 
to explain the $\gamma$-ray data. Such a number is compatible with the 
estimate of the compact remnants in the very central region around the GC 
\cite{2000ApJ...545..847M,2007MNRAS.377..897D}.

We further investigate the spatial distribution of the bulge MSP population,
using LMXBs as tracers. Assuming a spatial density profile of $r^{-2.4}$
we can well reproduce the observed LMXB distribution within $10^{\circ}$
around the GC \cite{2008A&A...491..209R}. Such a density profile is
quite consistent with that required to explain the GC $\gamma$-ray excess.
However, we still need to keep in mind that the current constraint on
the number density profile of LMXBs in the GC region is poor. It is 
possible that the density profile of LMXBs is slightly elongated along the 
Galactic plane as expected from the stellar model. In that case the MSP
scenario may have some tension with the $\gamma$-ray data
\cite{2014arXiv1402.6703D}.

We show in this work that the MSP population can naturally explain
the $\gamma$-ray excess in the GC region. It should be pointed out that
any other astrophysical populations with similar spectral, luminosity and 
spatial characteristics as the MSPs could also be the origin of the excess.
In any case, MSPs are the most natural sources to satisfy these constraints.

We note that some analyses claimed the $\gamma$-ray excesses 
extend to even larger scales in the inner Galaxy \cite{2013PDU.....2..118H,
2013arXiv1307.6862H,2014arXiv1402.6703D}. The excess spectra in these
regions seem to be even harder than that in the GC, and may be difficult
to be explained by MSPs \cite{2013PhRvD..88h3009H}. However, the
analysis at large scales may suffer from uncertainties from the large
scale diffuse background subtraction, especially if the emission from the
Fermi bubbles is not uniform \cite{2014arXiv1402.0403Y}. In spite that 
there are also uncertainties from the diffuse backgrounds, the results from
the GC analysis seem to be more robust \cite{2013PhRvD..88h3521G,
2014PhRvD..89f3515M}. Nevertheless, if the $\gamma$-ray excess does
extend to larger scales ($\gg 10^{\circ}$ from GC), the MSP scenario may
face difficulty.

Finally we propose that multi-wavelength observations of the counterpart
of the $\gamma$-ray excess, in e.g. X-rays, may help verify its existence 
as well as identify its nature. The X-ray emission from the MSPs and 
possibly the binary systems may show different properties (flux, skymap 
and spectrum) compared with that from DM annihilation, which could be
detectable by e.g., NuSTAR and other future X-ray missions.

\section*{Acknowledgments}
We thank X.-J. Bi, Y.-Z. Fan, A. Harding, L. Ho, D. Hooper, T. Linden, 
T. P. H. Tam and W. Wang for valuable comments and discussion. 
This work is supported by 973 Program under Grant No. 2013CB837000, and
by National Natural Science Foundation of China under Grant No. 11105155
(for QY).

\bibliographystyle{apsrev}
\bibliography{/home/yuanq/work/cygnus/tex/refs}

\begin{thebibliography}{51}
\expandafter\ifx\csname natexlab\endcsname\relax\def\natexlab#1{#1}\fi
\expandafter\ifx\csname bibnamefont\endcsname\relax
  \def\bibnamefont#1{#1}\fi
\expandafter\ifx\csname bibfnamefont\endcsname\relax
  \def\bibfnamefont#1{#1}\fi
\expandafter\ifx\csname citenamefont\endcsname\relax
  \def\citenamefont#1{#1}\fi
\expandafter\ifx\csname url\endcsname\relax
  \def\url#1{\texttt{#1}}\fi
\expandafter\ifx\csname urlprefix\endcsname\relax\def\urlprefix{URL }\fi
\providecommand{\bibinfo}[2]{#2}
\providecommand{\eprint}[2][]{\url{#2}}

\bibitem[{\citenamefont{{Goodenough} and {Hooper}}(2009)}]{2009arXiv0910.2998G}
\bibinfo{author}{\bibfnamefont{L.}~\bibnamefont{{Goodenough}}}
  \bibnamefont{and} \bibinfo{author}{\bibfnamefont{D.}~\bibnamefont{{Hooper}}},
  \bibinfo{journal}{ArXiv e-prints:0910.2998}  (\bibinfo{year}{2009}),
  \eprint{0910.2998}.

\bibitem[{\citenamefont{{Vitale} et~al.}(2009)\citenamefont{{Vitale},
  {Morselli}, and {for the Fermi/LAT Collaboration}}}]{2009arXiv0912.3828V}
\bibinfo{author}{\bibfnamefont{V.}~\bibnamefont{{Vitale}}},
  \bibinfo{author}{\bibfnamefont{A.}~\bibnamefont{{Morselli}}},
  \bibnamefont{and} \bibinfo{author}{\bibnamefont{{for the Fermi/LAT
  Collaboration}}}, \bibinfo{journal}{ArXiv e-prints:0912.3828}
  (\bibinfo{year}{2009}), \eprint{0912.3828}.

\bibitem[{\citenamefont{{Hooper} and {Goodenough}}(2011)}]{2011PhLB..697..412H}
\bibinfo{author}{\bibfnamefont{D.}~\bibnamefont{{Hooper}}} \bibnamefont{and}
  \bibinfo{author}{\bibfnamefont{L.}~\bibnamefont{{Goodenough}}},
  \bibinfo{journal}{Physics Letters B} \textbf{\bibinfo{volume}{697}},
  \bibinfo{pages}{412} (\bibinfo{year}{2011}), \eprint{1010.2752}.

\bibitem[{\citenamefont{{Boyarsky} et~al.}(2011)\citenamefont{{Boyarsky},
  {Malyshev}, and {Ruchayskiy}}}]{2011PhLB..705..165B}
\bibinfo{author}{\bibfnamefont{A.}~\bibnamefont{{Boyarsky}}},
  \bibinfo{author}{\bibfnamefont{D.}~\bibnamefont{{Malyshev}}},
  \bibnamefont{and}
  \bibinfo{author}{\bibfnamefont{O.}~\bibnamefont{{Ruchayskiy}}},
  \bibinfo{journal}{Physics Letters B} \textbf{\bibinfo{volume}{705}},
  \bibinfo{pages}{165} (\bibinfo{year}{2011}), \eprint{1012.5839}.

\bibitem[{\citenamefont{{Abazajian} and
  {Kaplinghat}}(2012)}]{2012PhRvD..86h3511A}
\bibinfo{author}{\bibfnamefont{K.~N.} \bibnamefont{{Abazajian}}}
  \bibnamefont{and}
  \bibinfo{author}{\bibfnamefont{M.}~\bibnamefont{{Kaplinghat}}},
  \bibinfo{journal}{\prd} \textbf{\bibinfo{volume}{86}}, \bibinfo{eid}{083511}
  (\bibinfo{year}{2012}), \eprint{1207.6047}.

\bibitem[{\citenamefont{{Gordon} and
  {Mac{\'{\i}}as}}(2013)}]{2013PhRvD..88h3521G}
\bibinfo{author}{\bibfnamefont{C.}~\bibnamefont{{Gordon}}} \bibnamefont{and}
  \bibinfo{author}{\bibfnamefont{O.}~\bibnamefont{{Mac{\'{\i}}as}}},
  \bibinfo{journal}{\prd} \textbf{\bibinfo{volume}{88}}, \bibinfo{eid}{083521}
  (\bibinfo{year}{2013}), \eprint{1306.5725}.

\bibitem[{\citenamefont{{Abazajian} et~al.}(2014)\citenamefont{{Abazajian},
  {Canac}, {Horiuchi}, and {Kaplinghat}}}]{2014arXiv1402.4090A}
\bibinfo{author}{\bibfnamefont{K.~N.} \bibnamefont{{Abazajian}}},
  \bibinfo{author}{\bibfnamefont{N.}~\bibnamefont{{Canac}}},
  \bibinfo{author}{\bibfnamefont{S.}~\bibnamefont{{Horiuchi}}},
  \bibnamefont{and}
  \bibinfo{author}{\bibfnamefont{M.}~\bibnamefont{{Kaplinghat}}},
  \bibinfo{journal}{ArXiv e-prints:1402.4090}  (\bibinfo{year}{2014}),
  \eprint{1402.4090}.

\bibitem[{\citenamefont{{Daylan} et~al.}(2014)\citenamefont{{Daylan},
  {Finkbeiner}, {Hooper}, {Linden}, {Portillo}, {Rodd}, and
  {Slatyer}}}]{2014arXiv1402.6703D}
\bibinfo{author}{\bibfnamefont{T.}~\bibnamefont{{Daylan}}},
  \bibinfo{author}{\bibfnamefont{D.~P.} \bibnamefont{{Finkbeiner}}},
  \bibinfo{author}{\bibfnamefont{D.}~\bibnamefont{{Hooper}}},
  \bibinfo{author}{\bibfnamefont{T.}~\bibnamefont{{Linden}}},
  \bibinfo{author}{\bibfnamefont{S.~K.~N.} \bibnamefont{{Portillo}}},
  \bibinfo{author}{\bibfnamefont{N.~L.} \bibnamefont{{Rodd}}},
  \bibnamefont{and} \bibinfo{author}{\bibfnamefont{T.~R.}
  \bibnamefont{{Slatyer}}}, \bibinfo{journal}{ArXiv e-prints:1402.6703}
  (\bibinfo{year}{2014}), \eprint{1402.6703}.

\bibitem[{\citenamefont{{Navarro} et~al.}(1997)\citenamefont{{Navarro},
  {Frenk}, and {White}}}]{1997ApJ...490..493N}
\bibinfo{author}{\bibfnamefont{J.~F.} \bibnamefont{{Navarro}}},
  \bibinfo{author}{\bibfnamefont{C.~S.} \bibnamefont{{Frenk}}},
  \bibnamefont{and} \bibinfo{author}{\bibfnamefont{S.~D.~M.}
  \bibnamefont{{White}}}, \bibinfo{journal}{\apj}
  \textbf{\bibinfo{volume}{490}}, \bibinfo{pages}{493} (\bibinfo{year}{1997}),
  \eprint{astro-ph/9611107}.

\bibitem[{\citenamefont{{Zhao}}(1996)}]{1996MNRAS.278..488Z}
\bibinfo{author}{\bibfnamefont{H.}~\bibnamefont{{Zhao}}},
  \bibinfo{journal}{\mnras} \textbf{\bibinfo{volume}{278}},
  \bibinfo{pages}{488} (\bibinfo{year}{1996}), \eprint{astro-ph/9509122}.

\bibitem[{\citenamefont{{Su} et~al.}(2010)\citenamefont{{Su}, {Slatyer}, and
  {Finkbeiner}}}]{2010ApJ...724.1044S}
\bibinfo{author}{\bibfnamefont{M.}~\bibnamefont{{Su}}},
  \bibinfo{author}{\bibfnamefont{T.~R.} \bibnamefont{{Slatyer}}},
  \bibnamefont{and} \bibinfo{author}{\bibfnamefont{D.~P.}
  \bibnamefont{{Finkbeiner}}}, \bibinfo{journal}{\apj}
  \textbf{\bibinfo{volume}{724}}, \bibinfo{pages}{1044} (\bibinfo{year}{2010}),
  \eprint{1005.5480}.

\bibitem[{\citenamefont{{Hooper} and {Slatyer}}(2013)}]{2013PDU.....2..118H}
\bibinfo{author}{\bibfnamefont{D.}~\bibnamefont{{Hooper}}} \bibnamefont{and}
  \bibinfo{author}{\bibfnamefont{T.~R.} \bibnamefont{{Slatyer}}},
  \bibinfo{journal}{Physics of the Dark Universe} \textbf{\bibinfo{volume}{2}},
  \bibinfo{pages}{118} (\bibinfo{year}{2013}), \eprint{1302.6589}.

\bibitem[{\citenamefont{{Huang}
  et~al.}(2013{\natexlab{a}})\citenamefont{{Huang}, {Urbano}, and
  {Xue}}}]{2013arXiv1307.6862H}
\bibinfo{author}{\bibfnamefont{W.-C.} \bibnamefont{{Huang}}},
  \bibinfo{author}{\bibfnamefont{A.}~\bibnamefont{{Urbano}}}, \bibnamefont{and}
  \bibinfo{author}{\bibfnamefont{W.}~\bibnamefont{{Xue}}},
  \bibinfo{journal}{ArXiv e-prints:1307.6862}
  (\bibinfo{year}{2013}{\natexlab{a}}), \eprint{1307.6862}.

\bibitem[{\citenamefont{{Hooper} and {Linden}}(2011)}]{2011PhRvD..84l3005H}
\bibinfo{author}{\bibfnamefont{D.}~\bibnamefont{{Hooper}}} \bibnamefont{and}
  \bibinfo{author}{\bibfnamefont{T.}~\bibnamefont{{Linden}}},
  \bibinfo{journal}{\prd} \textbf{\bibinfo{volume}{84}}, \bibinfo{eid}{123005}
  (\bibinfo{year}{2011}), \eprint{1110.0006}.

\bibitem[{\citenamefont{{Marshall} and
  {Primulando}}(2011)}]{2011JHEP...05..026M}
\bibinfo{author}{\bibfnamefont{G.}~\bibnamefont{{Marshall}}} \bibnamefont{and}
  \bibinfo{author}{\bibfnamefont{R.}~\bibnamefont{{Primulando}}},
  \bibinfo{journal}{Journal of High Energy Physics}
  \textbf{\bibinfo{volume}{5}}, \bibinfo{pages}{26} (\bibinfo{year}{2011}),
  \eprint{1102.0492}.

\bibitem[{\citenamefont{{Zhu}}(2011)}]{2011PhRvD..83g6011Z}
\bibinfo{author}{\bibfnamefont{G.}~\bibnamefont{{Zhu}}},
  \bibinfo{journal}{\prd} \textbf{\bibinfo{volume}{83}}, \bibinfo{eid}{076011}
  (\bibinfo{year}{2011}), \eprint{1101.4387}.

\bibitem[{\citenamefont{{Huang}
  et~al.}(2013{\natexlab{b}})\citenamefont{{Huang}, {Urbano}, and
  {Xue}}}]{2013arXiv1310.7609H}
\bibinfo{author}{\bibfnamefont{W.-C.} \bibnamefont{{Huang}}},
  \bibinfo{author}{\bibfnamefont{A.}~\bibnamefont{{Urbano}}}, \bibnamefont{and}
  \bibinfo{author}{\bibfnamefont{W.}~\bibnamefont{{Xue}}},
  \bibinfo{journal}{ArXiv e-prints:1310.7609}
  (\bibinfo{year}{2013}{\natexlab{b}}), \eprint{1310.7609}.

\bibitem[{\citenamefont{{Prasad Modak} et~al.}(2013)\citenamefont{{Prasad
  Modak}, {Majumdar}, and {Rakshit}}}]{2013arXiv1312.7488P}
\bibinfo{author}{\bibfnamefont{K.}~\bibnamefont{{Prasad Modak}}},
  \bibinfo{author}{\bibfnamefont{D.}~\bibnamefont{{Majumdar}}},
  \bibnamefont{and}
  \bibinfo{author}{\bibfnamefont{S.}~\bibnamefont{{Rakshit}}},
  \bibinfo{journal}{ArXiv e-prints}  (\bibinfo{year}{2013}),
  \eprint{1312.7488}.

\bibitem[{\citenamefont{{Boehm} et~al.}(2014)\citenamefont{{Boehm}, {Dolan},
  {McCabe}, {Spannowsky}, and {Wallace}}}]{2014arXiv1401.6458B}
\bibinfo{author}{\bibfnamefont{C.}~\bibnamefont{{Boehm}}},
  \bibinfo{author}{\bibfnamefont{M.~J.} \bibnamefont{{Dolan}}},
  \bibinfo{author}{\bibfnamefont{C.}~\bibnamefont{{McCabe}}},
  \bibinfo{author}{\bibfnamefont{M.}~\bibnamefont{{Spannowsky}}},
  \bibnamefont{and} \bibinfo{author}{\bibfnamefont{C.~J.}
  \bibnamefont{{Wallace}}}, \bibinfo{journal}{ArXiv e-prints:1401.6458}
  (\bibinfo{year}{2014}), \eprint{1401.6458}.

\bibitem[{\citenamefont{{Lacroix} et~al.}(2014)\citenamefont{{Lacroix},
  {Boehm}, and {Silk}}}]{2014arXiv1403.1987L}
\bibinfo{author}{\bibfnamefont{T.}~\bibnamefont{{Lacroix}}},
  \bibinfo{author}{\bibfnamefont{C.}~\bibnamefont{{Boehm}}}, \bibnamefont{and}
  \bibinfo{author}{\bibfnamefont{J.}~\bibnamefont{{Silk}}},
  \bibinfo{journal}{ArXiv e-prints:1403.1987}  (\bibinfo{year}{2014}),
  \eprint{1403.1987}.

\bibitem[{\citenamefont{{Agrawal} et~al.}(2014)\citenamefont{{Agrawal},
  {Batell}, {Hooper}, and {Lin}}}]{2014arXiv1404.1373A}
\bibinfo{author}{\bibfnamefont{P.}~\bibnamefont{{Agrawal}}},
  \bibinfo{author}{\bibfnamefont{B.}~\bibnamefont{{Batell}}},
  \bibinfo{author}{\bibfnamefont{D.}~\bibnamefont{{Hooper}}}, \bibnamefont{and}
  \bibinfo{author}{\bibfnamefont{T.}~\bibnamefont{{Lin}}},
  \bibinfo{journal}{ArXiv e-prints:1404.1373}  (\bibinfo{year}{2014}),
  \eprint{1404.1373}.

\bibitem[{\citenamefont{{Abazajian}}(2011)}]{2011JCAP...03..010A}
\bibinfo{author}{\bibfnamefont{K.~N.} \bibnamefont{{Abazajian}}},
  \bibinfo{journal}{\jcap} \textbf{\bibinfo{volume}{3}}, \bibinfo{eid}{010}
  (\bibinfo{year}{2011}), \eprint{1011.4275}.

\bibitem[{\citenamefont{{Mirabal}}(2013)}]{2013MNRAS.436.2461M}
\bibinfo{author}{\bibfnamefont{N.}~\bibnamefont{{Mirabal}}},
  \bibinfo{journal}{\mnras} \textbf{\bibinfo{volume}{436}},
  \bibinfo{pages}{2461} (\bibinfo{year}{2013}), \eprint{1309.3428}.

\bibitem[{\citenamefont{{Wang} et~al.}(2005)\citenamefont{{Wang}, {Jiang}, and
  {Cheng}}}]{2005MNRAS.358..263W}
\bibinfo{author}{\bibfnamefont{W.}~\bibnamefont{{Wang}}},
  \bibinfo{author}{\bibfnamefont{Z.~J.} \bibnamefont{{Jiang}}},
  \bibnamefont{and} \bibinfo{author}{\bibfnamefont{K.~S.}
  \bibnamefont{{Cheng}}}, \bibinfo{journal}{\mnras}
  \textbf{\bibinfo{volume}{358}}, \bibinfo{pages}{263} (\bibinfo{year}{2005}),
  \eprint{astro-ph/0501245}.

\bibitem[{\citenamefont{{Aprile} et~al.}(2012)\citenamefont{{Aprile},
  {Alfonsi}, {Arisaka}, {Arneodo}, {Balan}, {Baudis}, {Bauermeister},
  {Behrens}, {Beltrame}, {Bokeloh} et~al.}}]{2012PhRvL.109r1301A}
\bibinfo{author}{\bibfnamefont{E.}~\bibnamefont{{Aprile}}},
  \bibnamefont{et~al.}, \bibinfo{journal}{Physical Review Letters}
  \textbf{\bibinfo{volume}{109}}, \bibinfo{eid}{181301} (\bibinfo{year}{2012}),
  \eprint{1207.5988}.

\bibitem[{\citenamefont{{Akerib} et~al.}(2014)\citenamefont{{Akerib},
  {Ara{\'u}jo}, {Bai}, {Bailey}, {Balajthy}, {Bedikian}, {Bernard},
  {Bernstein}, {Bolozdynya}, {Bradley} et~al.}}]{2014PhRvL.112i1303A}
\bibinfo{author}{\bibfnamefont{D.~S.} \bibnamefont{{Akerib}}},
  \bibnamefont{et~al.}, \bibinfo{journal}{Physical Review Letters}
  \textbf{\bibinfo{volume}{112}}, \bibinfo{eid}{091303} (\bibinfo{year}{2014}),
  \eprint{1310.8214}.

\bibitem[{\citenamefont{{Macias} and {Gordon}}(2014)}]{2014PhRvD..89f3515M}
\bibinfo{author}{\bibfnamefont{O.}~\bibnamefont{{Macias}}} \bibnamefont{and}
  \bibinfo{author}{\bibfnamefont{C.}~\bibnamefont{{Gordon}}},
  \bibinfo{journal}{\prd} \textbf{\bibinfo{volume}{89}}, \bibinfo{eid}{063515}
  (\bibinfo{year}{2014}), \eprint{1312.6671}.

\bibitem[{\citenamefont{{Abdo} et~al.}(2013)\citenamefont{{Abdo}, {Ajello},
  {Allafort}, {Baldini}, {Ballet}, {Barbiellini}, {Baring}, {Bastieri},
  {Belfiore}, {Bellazzini} et~al.}}]{2013ApJS..208...17A}
\bibinfo{author}{\bibfnamefont{A.~A.} \bibnamefont{{Abdo}}},
  \bibnamefont{et~al.}, \bibinfo{journal}{\apjs}
  \textbf{\bibinfo{volume}{208}}, \bibinfo{eid}{17} (\bibinfo{year}{2013}),
  \eprint{1305.4385}.

\bibitem[{\citenamefont{{Abdo} et~al.}(2010)\citenamefont{{Abdo}, {Ackermann},
  {Ajello}, {Baldini}, {Ballet}, {Barbiellini}, {Bastieri}, {Bellazzini},
  {Blandford}, {Bloom} et~al.}}]{2010A&A...524A..75A}
\bibinfo{author}{\bibfnamefont{A.~A.} \bibnamefont{{Abdo}}},
  \bibnamefont{et~al.}, \bibinfo{journal}{\aap} \textbf{\bibinfo{volume}{524}},
  \bibinfo{pages}{A75} (\bibinfo{year}{2010}).

\bibitem[{\citenamefont{{Gnedin} et~al.}(2004)\citenamefont{{Gnedin},
  {Kravtsov}, {Klypin}, and {Nagai}}}]{2004ApJ...616...16G}
\bibinfo{author}{\bibfnamefont{O.~Y.} \bibnamefont{{Gnedin}}},
  \bibinfo{author}{\bibfnamefont{A.~V.} \bibnamefont{{Kravtsov}}},
  \bibinfo{author}{\bibfnamefont{A.~A.} \bibnamefont{{Klypin}}},
  \bibnamefont{and} \bibinfo{author}{\bibfnamefont{D.}~\bibnamefont{{Nagai}}},
  \bibinfo{journal}{\apj} \textbf{\bibinfo{volume}{616}}, \bibinfo{pages}{16}
  (\bibinfo{year}{2004}), \eprint{astro-ph/0406247}.

\bibitem[{\citenamefont{{Gnedin} et~al.}(2011)\citenamefont{{Gnedin},
  {Ceverino}, {Gnedin}, {Klypin}, {Kravtsov}, {Levine}, {Nagai}, and
  {Yepes}}}]{2011arXiv1108.5736G}
\bibinfo{author}{\bibfnamefont{O.~Y.} \bibnamefont{{Gnedin}}},
  \bibinfo{author}{\bibfnamefont{D.}~\bibnamefont{{Ceverino}}},
  \bibinfo{author}{\bibfnamefont{N.~Y.} \bibnamefont{{Gnedin}}},
  \bibinfo{author}{\bibfnamefont{A.~A.} \bibnamefont{{Klypin}}},
  \bibinfo{author}{\bibfnamefont{A.~V.} \bibnamefont{{Kravtsov}}},
  \bibinfo{author}{\bibfnamefont{R.}~\bibnamefont{{Levine}}},
  \bibinfo{author}{\bibfnamefont{D.}~\bibnamefont{{Nagai}}}, \bibnamefont{and}
  \bibinfo{author}{\bibfnamefont{G.}~\bibnamefont{{Yepes}}},
  \bibinfo{journal}{ArXiv e-prints:1108.5736}  (\bibinfo{year}{2011}),
  \eprint{1108.5736}.

\bibitem[{\citenamefont{{Voss} and
  {Gilfanov}}(2007{\natexlab{a}})}]{2007A&A...468...49V}
\bibinfo{author}{\bibfnamefont{R.}~\bibnamefont{{Voss}}} \bibnamefont{and}
  \bibinfo{author}{\bibfnamefont{M.}~\bibnamefont{{Gilfanov}}},
  \bibinfo{journal}{\aap} \textbf{\bibinfo{volume}{468}}, \bibinfo{pages}{49}
  (\bibinfo{year}{2007}{\natexlab{a}}), \eprint{astro-ph/0610649}.

\bibitem[{\citenamefont{{Voss} and
  {Gilfanov}}(2007{\natexlab{b}})}]{2007MNRAS.380.1685V}
\bibinfo{author}{\bibfnamefont{R.}~\bibnamefont{{Voss}}} \bibnamefont{and}
  \bibinfo{author}{\bibfnamefont{M.}~\bibnamefont{{Gilfanov}}},
  \bibinfo{journal}{\mnras} \textbf{\bibinfo{volume}{380}},
  \bibinfo{pages}{1685} (\bibinfo{year}{2007}{\natexlab{b}}),
  \eprint{astro-ph/0702580}.

\bibitem[{\citenamefont{{Hooper} et~al.}(2013)\citenamefont{{Hooper}, {Cholis},
  {Linden}, {Siegal-Gaskins}, and {Slatyer}}}]{2013PhRvD..88h3009H}
\bibinfo{author}{\bibfnamefont{D.}~\bibnamefont{{Hooper}}},
  \bibinfo{author}{\bibfnamefont{I.}~\bibnamefont{{Cholis}}},
  \bibinfo{author}{\bibfnamefont{T.}~\bibnamefont{{Linden}}},
  \bibinfo{author}{\bibfnamefont{J.~M.} \bibnamefont{{Siegal-Gaskins}}},
  \bibnamefont{and} \bibinfo{author}{\bibfnamefont{T.~R.}
  \bibnamefont{{Slatyer}}}, \bibinfo{journal}{\prd}
  \textbf{\bibinfo{volume}{88}}, \bibinfo{eid}{083009} (\bibinfo{year}{2013}),
  \eprint{1305.0830}.

\bibitem[{\citenamefont{{Faucher-Gigu{\`e}re} and
  {Loeb}}(2010)}]{2010JCAP...01..005F}
\bibinfo{author}{\bibfnamefont{C.-A.} \bibnamefont{{Faucher-Gigu{\`e}re}}}
  \bibnamefont{and} \bibinfo{author}{\bibfnamefont{A.}~\bibnamefont{{Loeb}}},
  \bibinfo{journal}{\jcap} \textbf{\bibinfo{volume}{1}}, \bibinfo{eid}{005}
  (\bibinfo{year}{2010}), \eprint{0904.3102}.

\bibitem[{\citenamefont{{Atwood} et~al.}(2009)\citenamefont{{Atwood}, {Abdo},
  {Ackermann}, {Althouse}, {Anderson}, {Axelsson}, {Baldini}, {Ballet}, {Band},
  {Barbiellini} et~al.}}]{2009ApJ...697.1071A}
\bibinfo{author}{\bibfnamefont{W.~B.} \bibnamefont{{Atwood}}},
  \bibnamefont{et~al.}, \bibinfo{journal}{\apj} \textbf{\bibinfo{volume}{697}},
  \bibinfo{pages}{1071} (\bibinfo{year}{2009}), \eprint{0902.1089}.

\bibitem[{\citenamefont{{Daugherty} and {Harding}}(1996)}]{1996ApJ...458..278D}
\bibinfo{author}{\bibfnamefont{J.~K.} \bibnamefont{{Daugherty}}}
  \bibnamefont{and} \bibinfo{author}{\bibfnamefont{A.~K.}
  \bibnamefont{{Harding}}}, \bibinfo{journal}{\apj}
  \textbf{\bibinfo{volume}{458}}, \bibinfo{pages}{278} (\bibinfo{year}{1996}),
  \eprint{astro-ph/9508155}.

\bibitem[{\citenamefont{{Muslimov} and
  {Harding}}(2004{\natexlab{a}})}]{2004ApJ...617..471M}
\bibinfo{author}{\bibfnamefont{A.~G.} \bibnamefont{{Muslimov}}}
  \bibnamefont{and} \bibinfo{author}{\bibfnamefont{A.~K.}
  \bibnamefont{{Harding}}}, \bibinfo{journal}{\apj}
  \textbf{\bibinfo{volume}{617}}, \bibinfo{pages}{471}
  (\bibinfo{year}{2004}{\natexlab{a}}), \eprint{astro-ph/0408377}.

\bibitem[{\citenamefont{{Cheng} et~al.}(1986)\citenamefont{{Cheng}, {Ho}, and
  {Ruderman}}}]{1986ApJ...300..500C}
\bibinfo{author}{\bibfnamefont{K.~S.} \bibnamefont{{Cheng}}},
  \bibinfo{author}{\bibfnamefont{C.}~\bibnamefont{{Ho}}}, \bibnamefont{and}
  \bibinfo{author}{\bibfnamefont{M.}~\bibnamefont{{Ruderman}}},
  \bibinfo{journal}{\apj} \textbf{\bibinfo{volume}{300}}, \bibinfo{pages}{500}
  (\bibinfo{year}{1986}).

\bibitem[{\citenamefont{{Muslimov} and
  {Harding}}(2004{\natexlab{b}})}]{2004ApJ...606.1143M}
\bibinfo{author}{\bibfnamefont{A.~G.} \bibnamefont{{Muslimov}}}
  \bibnamefont{and} \bibinfo{author}{\bibfnamefont{A.~K.}
  \bibnamefont{{Harding}}}, \bibinfo{journal}{\apj}
  \textbf{\bibinfo{volume}{606}}, \bibinfo{pages}{1143}
  (\bibinfo{year}{2004}{\natexlab{b}}), \eprint{astro-ph/0402462}.

\bibitem[{\citenamefont{{Dyks} and {Rudak}}(2003)}]{2003ApJ...598.1201D}
\bibinfo{author}{\bibfnamefont{J.}~\bibnamefont{{Dyks}}} \bibnamefont{and}
  \bibinfo{author}{\bibfnamefont{B.}~\bibnamefont{{Rudak}}},
  \bibinfo{journal}{\apj} \textbf{\bibinfo{volume}{598}}, \bibinfo{pages}{1201}
  (\bibinfo{year}{2003}), \eprint{astro-ph/0303006}.

\bibitem[{\citenamefont{{Qiao} et~al.}(2004)\citenamefont{{Qiao}, {Lee},
  {Wang}, {Xu}, and {Han}}}]{2004ApJ...606L..49Q}
\bibinfo{author}{\bibfnamefont{G.~J.} \bibnamefont{{Qiao}}},
  \bibinfo{author}{\bibfnamefont{K.~J.} \bibnamefont{{Lee}}},
  \bibinfo{author}{\bibfnamefont{H.~G.} \bibnamefont{{Wang}}},
  \bibinfo{author}{\bibfnamefont{R.~X.} \bibnamefont{{Xu}}}, \bibnamefont{and}
  \bibinfo{author}{\bibfnamefont{J.~L.} \bibnamefont{{Han}}},
  \bibinfo{journal}{\apjl} \textbf{\bibinfo{volume}{606}}, \bibinfo{pages}{L49}
  (\bibinfo{year}{2004}), \eprint{astro-ph/0403398}.

\bibitem[{\citenamefont{{Johnson} et~al.}(2014)\citenamefont{{Johnson},
  {Venter}, {Harding}, {Guillemot}, {Smith}, {Kramer}, {Celik}, {den Hartog},
  {Ferrara}, {Hou} et~al.}}]{2014arXiv1404.2264J}
\bibinfo{author}{\bibfnamefont{T.~J.} \bibnamefont{{Johnson}}},
  \bibnamefont{et~al.}, \bibinfo{journal}{ArXiv e-prints:1404.2264}
  (\bibinfo{year}{2014}), \eprint{1404.2264}.

\bibitem[{\citenamefont{{Harding} et~al.}(2002)\citenamefont{{Harding},
  {Muslimov}, and {Zhang}}}]{2002ApJ...576..366H}
\bibinfo{author}{\bibfnamefont{A.~K.} \bibnamefont{{Harding}}},
  \bibinfo{author}{\bibfnamefont{A.~G.} \bibnamefont{{Muslimov}}},
  \bibnamefont{and} \bibinfo{author}{\bibfnamefont{B.}~\bibnamefont{{Zhang}}},
  \bibinfo{journal}{\apj} \textbf{\bibinfo{volume}{576}}, \bibinfo{pages}{366}
  (\bibinfo{year}{2002}), \eprint{astro-ph/0205077}.

\bibitem[{\citenamefont{{Hirotani}}(2013)}]{2013ApJ...766...98H}
\bibinfo{author}{\bibfnamefont{K.}~\bibnamefont{{Hirotani}}},
  \bibinfo{journal}{\apj} \textbf{\bibinfo{volume}{766}}, \bibinfo{eid}{98}
  (\bibinfo{year}{2013}), \eprint{1301.5717}.

\bibitem[{\citenamefont{{Hessels} et~al.}(2006)\citenamefont{{Hessels},
  {Ransom}, {Stairs}, {Freire}, {Kaspi}, and {Camilo}}}]{2006Sci...311.1901H}
\bibinfo{author}{\bibfnamefont{J.~W.~T.} \bibnamefont{{Hessels}}},
  \bibinfo{author}{\bibfnamefont{S.~M.} \bibnamefont{{Ransom}}},
  \bibinfo{author}{\bibfnamefont{I.~H.} \bibnamefont{{Stairs}}},
  \bibinfo{author}{\bibfnamefont{P.~C.~C.} \bibnamefont{{Freire}}},
  \bibinfo{author}{\bibfnamefont{V.~M.} \bibnamefont{{Kaspi}}},
  \bibnamefont{and} \bibinfo{author}{\bibfnamefont{F.}~\bibnamefont{{Camilo}}},
  \bibinfo{journal}{Science} \textbf{\bibinfo{volume}{311}},
  \bibinfo{pages}{1901} (\bibinfo{year}{2006}), \eprint{astro-ph/0601337}.

\bibitem[{\citenamefont{{Miralda-Escud{\'e}} and
  {Gould}}(2000)}]{2000ApJ...545..847M}
\bibinfo{author}{\bibfnamefont{J.}~\bibnamefont{{Miralda-Escud{\'e}}}}
  \bibnamefont{and} \bibinfo{author}{\bibfnamefont{A.}~\bibnamefont{{Gould}}},
  \bibinfo{journal}{\apj} \textbf{\bibinfo{volume}{545}}, \bibinfo{pages}{847}
  (\bibinfo{year}{2000}), \eprint{astro-ph/0003269}.

\bibitem[{\citenamefont{{Deegan} and {Nayakshin}}(2007)}]{2007MNRAS.377..897D}
\bibinfo{author}{\bibfnamefont{P.}~\bibnamefont{{Deegan}}} \bibnamefont{and}
  \bibinfo{author}{\bibfnamefont{S.}~\bibnamefont{{Nayakshin}}},
  \bibinfo{journal}{\mnras} \textbf{\bibinfo{volume}{377}},
  \bibinfo{pages}{897} (\bibinfo{year}{2007}), \eprint{astro-ph/0611524}.

\bibitem[{\citenamefont{{Bhattacharya}}(1996)}]{1996ASPC..105..547B}
\bibinfo{author}{\bibfnamefont{D.}~\bibnamefont{{Bhattacharya}}}, in
  \emph{\bibinfo{booktitle}{IAU Colloq. 160: Pulsars: Problems and Progress}},
  edited by \bibinfo{editor}{\bibfnamefont{S.}~\bibnamefont{{Johnston}}},
  \bibinfo{editor}{\bibfnamefont{M.~A.} \bibnamefont{{Walker}}},
  \bibnamefont{and} \bibinfo{editor}{\bibfnamefont{M.}~\bibnamefont{{Bailes}}}
  (\bibinfo{year}{1996}), vol. \bibinfo{volume}{105} of
  \emph{\bibinfo{series}{Astronomical Society of the Pacific Conference
  Series}}, p. \bibinfo{pages}{547}.

\bibitem[{\citenamefont{{Revnivtsev} et~al.}(2008)\citenamefont{{Revnivtsev},
  {Lutovinov}, {Churazov}, {Sazonov}, {Gilfanov}, {Grebenev}, and
  {Sunyaev}}}]{2008A&A...491..209R}
\bibinfo{author}{\bibfnamefont{M.}~\bibnamefont{{Revnivtsev}}},
  \bibinfo{author}{\bibfnamefont{A.}~\bibnamefont{{Lutovinov}}},
  \bibinfo{author}{\bibfnamefont{E.}~\bibnamefont{{Churazov}}},
  \bibinfo{author}{\bibfnamefont{S.}~\bibnamefont{{Sazonov}}},
  \bibinfo{author}{\bibfnamefont{M.}~\bibnamefont{{Gilfanov}}},
  \bibinfo{author}{\bibfnamefont{S.}~\bibnamefont{{Grebenev}}},
  \bibnamefont{and}
  \bibinfo{author}{\bibfnamefont{R.}~\bibnamefont{{Sunyaev}}},
  \bibinfo{journal}{\aap} \textbf{\bibinfo{volume}{491}}, \bibinfo{pages}{209}
  (\bibinfo{year}{2008}), \eprint{0805.0259}.

\bibitem[{\citenamefont{{Yang} et~al.}(2014)\citenamefont{{Yang}, {Aharonian},
  and {Crocker}}}]{2014arXiv1402.0403Y}
\bibinfo{author}{\bibfnamefont{R.-z.} \bibnamefont{{Yang}}},
  \bibinfo{author}{\bibfnamefont{F.}~\bibnamefont{{Aharonian}}},
  \bibnamefont{and}
  \bibinfo{author}{\bibfnamefont{R.}~\bibnamefont{{Crocker}}},
  \bibinfo{journal}{ArXiv e-prints:1402.0403}  (\bibinfo{year}{2014}),
  \eprint{1402.0403}.

\end{thebibliography}

\end{document}